\documentclass[apj]{emulateapj}
\usepackage{color}
\usepackage{natbib}
\usepackage{apjfonts}
\usepackage{appendix}
\definecolor{kaki}{cmyk}{0,0.2,1,0}

\def\lsim{\lower 2pt \hbox{$\, \buildrel {\scriptstyle <}\over
{\scriptstyle \sim}\,$}}
\def\gsim{\lower 2pt \hbox{$\, \buildrel {\scriptstyle >}\over
{\scriptstyle \sim}\,$}}

\shorttitle{Chromospheric and Coronal Wave Generation}
\shortauthors{Yoshiaki Kato et al.}

\begin{document}


\title{Chromospheric and Coronal Wave Generation in a Magnetic Flux Sheath}


\author{Yoshiaki Kato}
\affil{Institute of Theoretical Astrophysics, University of Oslo, P.O. Box 1029 Blindern, N-0315 Oslo, Norway}
\email{yoshiaki.kato@astro.uio.no}
\author{Oskar Steiner}
\affil{Kiepenheuer-Institut f\"ur Sonnenphysik, Sch\"oneckstrasse 6, D-79104 Freiburg, Germany}
\affil{Istituto Ricerche Solari Locarno (IRSOL), via Patocchi 57---Prato Pernice, 6605 Locarno-Monti, Switzerland}
%\email{steiner@kis.uni-freiburg.de}
%
\author{Viggo Hansteen}
%\email{viggo.hansteen@astro.uio.no}
\author{Boris Gudiksen}
\author{Sven Wedemeyer}
%\email{sven.wedemeyer-bohm@astro.uio.no}
\author{Mats Carlsson}
%\email{m.p.o.carlsson@astro.uio.no}
\affil{Institute of Theoretical Astrophysics, University of Oslo, P.O. Box 1029 Blindern, N-0315 Oslo, Norway}

\begin{abstract}
%% Here comes the abstract\\
%% \rule{1pt}{4cm}
Using radiation magnetohydrodynamic simulations of the solar atmospheric layers from the upper convection zone to the lower corona, we investigate the self-consistent excitation of slow magneto-acoustic body waves (slow modes) in a magnetic flux concentration.  We find that the convective downdrafts in the close surroundings of a two-dimensional flux slab ``pump'' the plasma inside it in the downward direction.  This action produces a downflow inside the flux slab, which encompasses ever higher layers, causing an upwardly propagating rarefaction wave.  The slow mode, excited by the adiabatic compression of the downflow near the optical surface, travels along the magnetic field in the upward direction at the tube speed.  It develops into a shock wave at chromospheric heights, where it dissipates, lifts the transition region, and produces an offspring in the form of a compressive wave that propagates further into the corona.  In the wake of downflows and propagating shock waves, the atmosphere inside the flux slab in the chromosphere and higher tends to oscillate with a period of $\nu\approx 4$~mHz.  We conclude that this process of ``magnetic pumping'' is a most plausible mechanism for the direct generation of longitudinal chromospheric and coronal compressive waves within magnetic flux concentrations, and it may provide an important heat source in the chromosphere.  It may also be responsible for certain types of dynamic fibrils. 
\end{abstract}


%%%%%%%%%%%%%%%%%%%%%%%%%%%%%%%%%%%%%%%%%%%%%%%%%%%%%%%%%%%%%%%%%%%%%%%%%%%%%%%%
\section{Introduction}
\label{sect1}
%%% Chromospheric emission and magnetic fields
The solar chromosphere is known to be a dynamic, structured medium.
Filtergrams in \ion{Ca}{2} H and K reveal two main sources of \ion{Ca}{2} emission: strong network patches and plage regions \citep{simon+leighton:1964}.  
Both coincide with small magnetic flux concentrations in the photosphere and with mottles/spicules in the chromosphere \citep{beckers:1968}.
The close relationship between \ion{Ca}{2} emission and magnetic flux \citep{skumanich+:1975,schrijver+:1989} suggests that the magnetic field plays a key role for the chromospheric emission and the formation of dynamic fibrils \citep{hansteen+:2006}.
The physical mechanism of this emission has remained elusive.
There are two plausible candidates, the dissipation of magnetohydrodynamic (MHD) waves and the direct dissipation of electric currents.
With regard to the former process, numerous studies have been carried out based on the approximation of slender flux tubes \citep[][and reference therein]{roberts+webb:1978,spruit:1981,hollweg+roberts:1981, hasan+ulmschneider:2004a}.
They have greatly expanded our understanding of the fundamental physics of MHD wave modes, mode coupling, dependency on the driving mechanism and the flux tube geometry, shock formation, etc.
Recent numerical simulations of the propagation of magneto-acoustic waves through an environment resembling small-scale magnetic flux concentrations with internal structure include those of \citet{hasan+van_Ballegooijen:2008}, \citet{khomenko+:2008}, \citet{vigeesh+:2009}, \citet{murawski+zaqarashvili:2010}, \citet{fedun+:2011}, \citet{vigeesh+:2012}, or \citet{mumford+:2015}.
\citet{murawski+:2015}, \citet{mumford+:2015}, and \citet{giagkiozis+:2015} performed numerical simulations of impulsively and monochromatically generated Alf\'ven waves in a solitary flux tube. 
All these models and simulations impose a given, arbitrary excitation, which is either monochromatic or impulsive, or derived from a theoretical spectrum of turbulence, or from observations.
%

%%% A foundational study
A pioneering study on the self-consistent excitation of MHD waves in a thick magnetic flux concentration was carried out by \citet{steiner+:1998}. 
They found (1) swaying motion of a flux sheet due to asymmetrical convective flow, (2) fast narrow downflows at the interface between the flux sheet and its environment in the subsurface layers, and (3) shocks inside and outside the flux sheet.
Recently, \citet{kato+:2011} found in their two-dimensional radiation magnetohydrodynamic simulations that longitudinal magneto-acoustic body waves inside the two-dimensional flux sheet are excited by a process which they termed ``magnetic pumping'' in which the downflow jets adjacent to the flux sheet initiate these waves.
\citet{heggland+:2011} also found similar waves that propagate upward along the magnetic flux sheet (see Figure\,2 of their paper), probably excited by this same pumping process.
Because their calculations reach all the way up to the lower corona, they found that these waves create what they call ``jets'' of various length and lifetimes once the waves reach the transition region.
%

%%% Our study
In this paper we investigate the self-consistent generation of MHD waves in a thick magnetic flux slab by means of a two-dimensional numerical simulation that takes radiative losses and heat conduction in the outer atmosphere into account, resulting in an interface region between the chromosphere and the corona.
The limitation to two spatial dimensions allows us to study oscillations in a long-lasting magnetic flux concentration without disturbance from instabilities that would likely arise in three dimensions.
The method and model we use in this study are described in  {\S{\ref{sect2}}}.
A comparison between theory and model regarding the generation and propagation of MHD waves is given in {\S{\ref{sect3}}}.
In \S{\ref{sect4}} we discuss the results and their limitations, and implications.
In \S{\ref{sect5}} we summarise our results.
%

%%%%%%%%%%%%%%%%%%%%%%%%%%%%%%%%%%%%%%%%%%%%%%%%%%%%%%%%%%%%%%%%%%%%%%%%%%%%%%%%
\section{Method and Model}
\label{sect2}

%%% About previous study
\citet{kato+:2011} carried out their simulations with the CO\raisebox{0.5ex}{\footnotesize 5}BOLD code \citep{freytag+:2012} in a box that reached to a height of $780$~km above the average height of $\tau_{500}=1$\footnote{Here and in the following, $z=0$ corresponds to the average optical depth $\tau_{500}=1$ in the region outside of the magnetic flux concentration.} in two-dimensional geometry.
In this study, we extend the computational box size up to the lower corona in order to diminish the impact of the top boundary on the flow and to include the propagation of waves across the transition region between the chromosphere and the corona.
For this purpose, we perform the simulation presented in this paper with the {\it Bifrost} code \citep{gudiksen+:2011} which is the ideal tool for investigating the generation and propagation of waves in the atmosphere from the upper convection zone to the lower corona.
%

%%% Present study
For the initial configuration, we start from a snapshot of the model atmosphere from the upper convection zone to the lower chromosphere of the previous calculation by \citet{kato+:2011}.  It already contains a single, isolated magnetic flux concentration that can be considered a slab of a flux sheath that is translational invariant in the direction perpendicular to the two-dimensional computational domain.
Next, we smoothly connect the thermodynamic state of it to a simple nearly hydrostatic model of the upper chromosphere, transition region and lower corona, based on the spatially averaged properties taken from a single snapshot of a two-dimensional {\it Bifrost} calculation.
Fixing the mean temperature at the upper boundary to $7\times 10^{5}$ K, the model is allowed to relax for a period of some 20 minutes simulation time, until transients propagate out of the computational box and a quasi-steady state is achieved in the upper atmosphere, in much the same way as was done by \citet{hansteen+:2006,de_pontieu+:2007a,heggland+:2011}.
After this state is achieved, the temperature gradient is set to zero at the top boundary, as this gives less wave reflection.
In the relaxation phase, the modeled corona cools to an average value of some $4\times 10^5$~K, after which it remains roughly at that temperature during the subsequent 68 minutes model run.
This model includes optically thick radiative losses in the photosphere and lower chromosphere, parameterized radiative losses in the upper chromosphere, optically thin radiation in the transition region and corona, and thermal conduction along the magnetic field.  Details of the physics included and solution methods employed may be found in \citet{gudiksen+:2011}.
The magnetic field in the region above the lower chromosphere is determined by an extrapolation of the magnetic field at the top boundary $z_{\rm 0}=780$~km of the previous study, using the Green-function method \citep{sakurai:1982}.
The extrapolated magnetic field is given by the vector potential
\begin{equation}
A(\mbox{\boldmath$r$})=\int B_{\rm z_{\rm 0}}(\mbox{\boldmath$r$}')G(\mbox{\boldmath$r$},\mbox{\boldmath$r$}'){\rm d}r',
\end{equation}
where $B_{\rm z_{\rm 0}}$ is the vertical component of the magnetic field at the top boundary of the previous study, and $G(\mbox{\boldmath$r$},\mbox{\boldmath$r$}')$ is the Green function for which we choose
\begin{equation}
G(\mbox{\boldmath$r$},\mbox{\boldmath$r$}')=\frac{1}{2\pi |\mbox{\boldmath$r$}-\mbox{\boldmath$r$}'|} + \exp{\left[-(z-z_{\rm 0})\right]} - 1,
\end{equation}
where $\mbox{\boldmath$r$}=(x,z)$ and $\mbox{\boldmath$r$}'=(x,z_{\rm 0})$.
By integrating over the computational domain above $z_{\rm 0}$, the extrapolated magnetic field smoothly connects to the magnetic field  beneath it, so that the magneto-acoustic waves can propagate without experiencing reflection or refraction as a result of discontinuities in the initial magnetic field.
In the following simulation, the magnetic field remains concentrated in an isolated magnetic flux concentration during more than 60~minutes real time.
The main purpose for investigating an isolated magnetic flux concentration is to extract the essential constituents of the dynamics of a single flux sheath rather than becoming distracted by the complex interactions between multiple flux concentrations such as merging and splitting.
%

%%% Figure 1: Snapshot of the simulation for explaining the model
\begin{figure*}
\epsscale{1.0}
\plotone{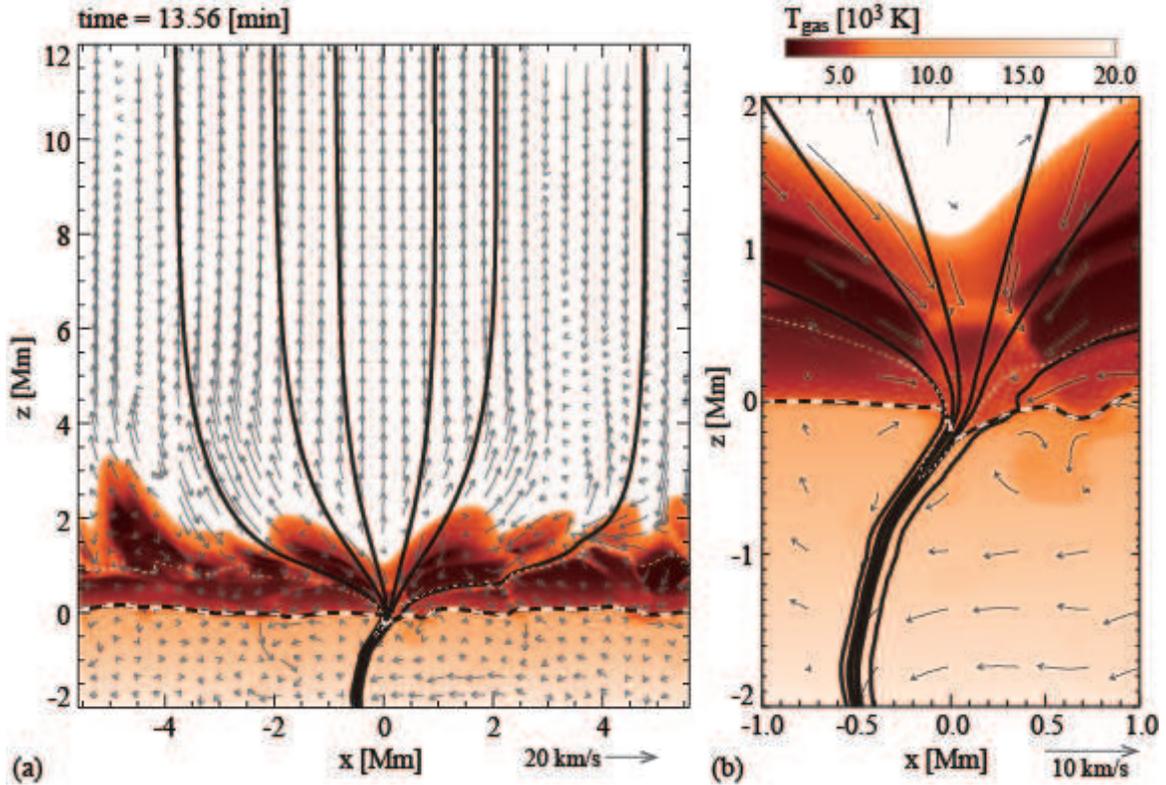}
%\plotone{f1grey.eps}
\caption{(a) Snapshot of the simulation, showing the full computational domain.  The domain ranges in the vertical direction from $z=-2.12\,{\rm Mm}$ to $z=11.96\,{\rm Mm}$, whereas the horizontal extension is, like in the previous study,  11.2\,{\rm Mm}.  Color scales show the gas temperature.  Arrows indicate the plasma velocity.  Black solid curves are representative magnetic field lines, which have components in the $x$-$z$-plane only.  The outer pair of field lines delineate the flux slab whose field strength is larger than $1$~kG at $z=0$.  It contains a magnetic flux of $\sim 10^{19}$~Mx when assuming a slab thickness of $100$~km perpendicular to the $x$-$z$-plane.  The middle pair of field lines delineate the ``body'' of the flux slab whose strength is larger than $1.5$~kG at $z=0$ (containing $40\%$ of the total magnetic flux).  The inner pair of field lines indicates the ``core'' of the flux slab containing $\sim 17\%$ of the total magnetic flux.  The horizontal width of the flux slab is roughly 200~km near the optical depth unity $\tau_{\rm 500} = 1$, which is indicated by the black and white dashed curve.  The white dotted curve shows the isosurface of plasma-$\beta$ unity, $\beta = 1$.  (b) Close-up of the lower part of the flux slab.  The arc-shaped temperature enhancement at $z\sim 700$ km in the core region of the flux slab arises from a propagating slow shock (see Sections\,\ref{sec:response} and Appendix \ref{sec:shocks} for details).}
\label{fig:snapshot}
\end{figure*}

%%% Describe Figure 1 (a)
Figure\,\ref{fig:snapshot}a shows a snapshot of the gas temperature in the entire computational domain, which extends over a height range of 14~Mm reaching 12~Mm above the mean surface of optical depth unity.
The model is computed on a numerical grid of $400\times 535$ cells which covers 11.2~Mm horizontally.
The grid-cell size in the horizontal direction is 28~km, in the vertical direction it is 13~km at $z=0$, continuously increasing to 30~km through the convection-zone and to 45~km at the upper boundary in the corona.
The lateral boundary conditions are periodic in all variables, whereas the lower boundary is open in the sense that the fluid can freely flow in and out of the computational domain.
Thereby, the total mass of the box remains close to constant.
The specific entropy of the inflowing mass is fixed to a value previously determined so as to yield the correct solar radiative flux at the upper boundary.
For the upper boundary we use the formulation of characteristics, which aims to transmit all disturbances with least reflections \citep[see][for more details]{gudiksen+:2011}.
%

%%% Describe Figure 1 (b)
Figure\,\ref{fig:snapshot}b shows a close-up of the lower part of the flux slab in panel (a).
The arc-shaped temperature enhancement at $z\sim 700$ km in the flux slab represents a propagating shock which resembles the time-sequence of a propagating shock reported by \citet{heggland+:2011} (see Figure\,2 of their paper) or by \citet{steiner+:1998} deeper down in the photosphere (see Figure\,1 of their paper).
The magnetic flux slab resides in a location of strong convective downdraft.
As a consequence of the interaction with the surrounding convective flow, the flux concentration moves laterally, gets distorted, and exhibits internal plasma flow.
This interaction excites magneto-acoustic waves within the flux concentration, in particular longitudinal slow modes which play an important role in the dynamics of it  as is shown in the next section.

%%%%%%%%%%%%%%%%%%%%%%%%%%%%%%%%%%%%%%%%%%%%%%%%%%%%%%%%%%%%%%%%%%%%%%%%%%%%%%%%
\section{Magnetic Pumping and the propagation of waves}
\label{sect3}
\begin{figure*}
\epsscale{1.0}
\plotone{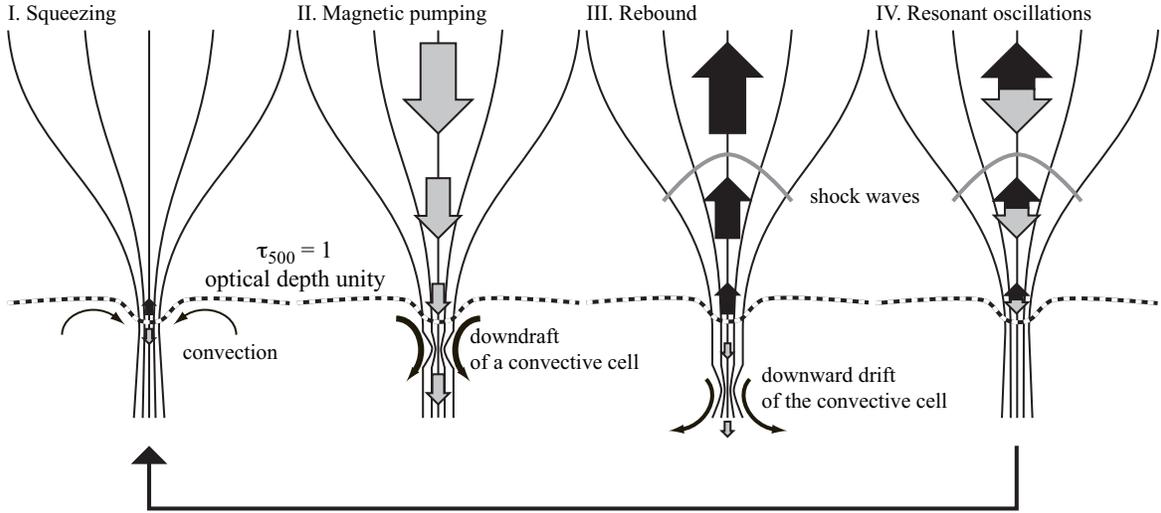}
\caption{Schematic picture of the magnetic pumping process.  Sporadic strong downdrafts in the close surroundings of the magnetic flux concentration, pump the plasma within the magnetic flux concentration in the downward direction by the action of inertial forces acting on the magnetic field.  As soon as the transient ambient downdraft weakens, the pumping comes to a halt and the downflowing plasma from the photospheric and chromospheric layers inside the magnetic flux concentration rebounds, which generates an upwardly propagating slow magneto-acoustic wave.  This process repeats every $5 - 15$ minutes, while in the meantime, the atmosphere inside the flux tube tends to oscillate at the cut-off period of $3 - 5$ minutes.}
\label{fig:schematic}
\end{figure*}

%%% Explain the basic understanding on magnetic pumping process
A novel mechanism of wave generation within small-scale magnetic flux concentrations called magnetic pumping, was described by \citet{kato+:2011}.
These authors discovered that sporadic strong downdrafts in the close surroundings of a magnetic flux concentration, pump plasma inside it in the downward direction by the action of inertial forces on the magnetic field -- a process that was first described by \citet{parker:1974a} who referred to it as turbulent pumping.
As soon as the transient ambient downdraft weakens, the pumping comes to a halt and the downflowing plasma from the photospheric and chromospheric layers of the magnetic element rebounds, which leads to an upwardly propagating slow magneto-acoustic wave.\footnote{This wave is of predominant acoustic nature, travels along the magnetic field in a low plasma-$\beta$ environment with about the speed of sound, and is therefore called a slow wave.}
It develops into a shock wave in the chromospheric layers of the magnetic element.
In the wake of a pumping event, the atmosphere of the flux concentration tends to oscillate at the cut-off period (see Figure\,\ref{fig:schematic} for illustration).
These pumping events are suspected to be a crucial mechanism for the excitation of longitudinal slow modes in magnetic flux concentrations, for the heating in network and plage areas, and possibly for the development of dynamic fibrils.
In the following subsections, we investigate details of this process and corresponding reactions at each layer of the atmosphere from the photosphere to the lower corona.
%

%%% Figure 2: Schematic picture of the magnetic pumping

\subsection{Overview of the flux-sheath atmosphere}
%%% Overview of the flux tube atmosphere in my simulation

%
The width of the flux slab is not more than 200\,km near $\tau_{\rm 500} = 1$ and rapidly expands with height to fill all of the lateral space above $z\approx 2$~Mm.
In the course of the simulation, the widths of the body and core of the flux slab vary in time by roughly $10\%$ of their time-averaged widths at all heights.
Above $\tau_{\rm 500} = 1$, where the width of the flux slab surpasses the pressure scale height, the thin flux-tube approximation \citep[e.g.,][]{spruit:1981,ferriz-mas+:1989} becomes questionable \cite[see also][]{steiner+pizzo:1989}.
In spite of this restriction, the linear analysis of the flux sheath above the convection zone might still provide us with useful insights regarding the nature of waves in that height range. 
The analytical solutions in Appendix\,\ref{sec:analytical_model} illustrate the well known fact that high-frequency disturbances ($\omega > \omega_{\rm v}$) can propagate along a flux tube/slab whereas low-frequency disturbances ($\omega < \omega_{\rm v}$) are evanescent.
Interesting in the present context is that in the wake of an impulsive disturbance, the atmosphere oscillates with the cut-off frequency $\omega_{\rm v}$, \citep[see, e.g.,][]{rae+roberts:1982,hasan+kalkofen:1999}.
In the following, we determine the atmospheric parameters in the core of the simulated flux slab and then evaluate the atmospheric responses such as the propagation speed of the tube wave and the characteristic frequencies of flux tube oscillations for better understanding the characteristic of waves and oscillations in the dynamic atmosphere.
%

%%% Describing time-averaged flux tube atmosphere in my simulation: Overview

%%% Figure 4: Atmospheric parameters
\begin{figure*}
\epsscale{1.0}
\plotone{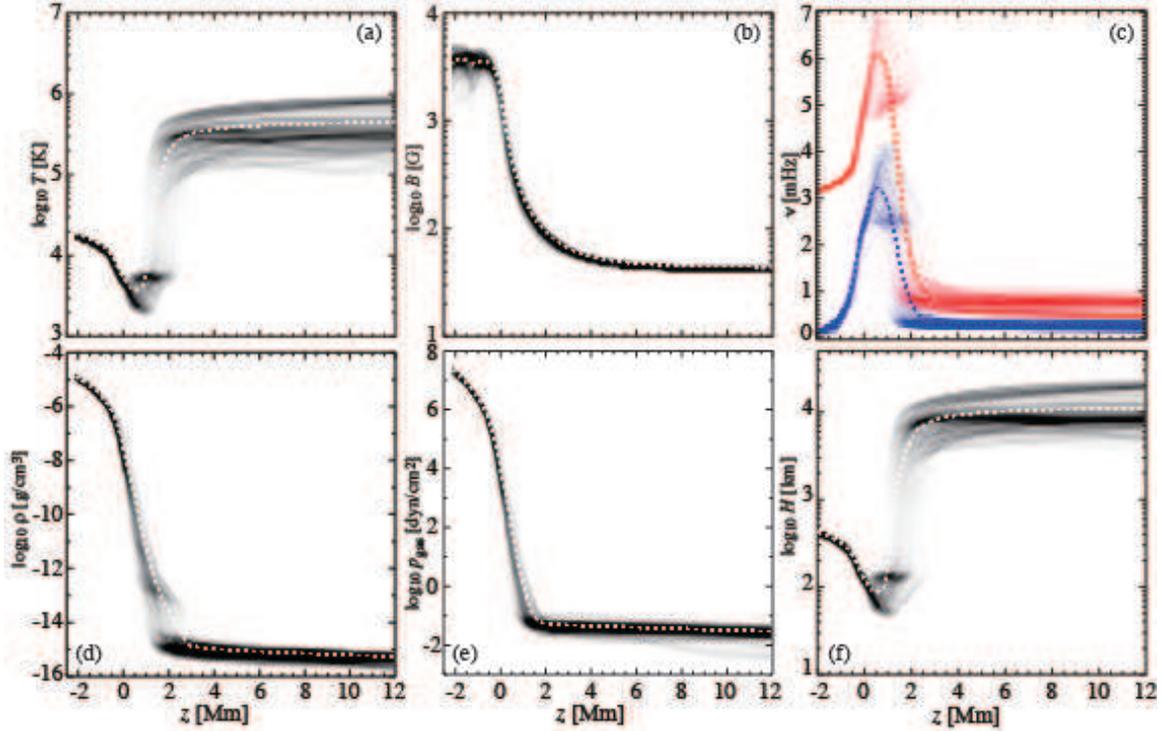}
\caption{Scatter plots of atmospheric parameters spatially averaged over the core of the flux slab as a function of height:  (a) temperature, (b) magnetic field strength, (c) cut-off frequencies for longitudinal tube/slab waves ($\omega_{\rm v}/2\pi$: red) and for transverse kink waves ($\omega_{\rm k}/2\pi$: blue), (d) gas density, (e) gas pressure, and (f) pressure scale height.  Dashed curves indicate time-averaged profiles of each atmospheric parameter.  Zero height corresponds to the average optical depth $\tau_{\rm 500}=1$ outside of the flux slab.  Height is counted positive in the upward direction (outward of the Sun).  The time-averaged Wilson depression is 200~km.}
\label{fig:fluxtube-profile}
\end{figure*}

In Figure\,\ref{fig:fluxtube-profile}, gray scales indicate variation in time and the white dotted curves show the time-averaged profiles of each atmospheric parameter.
The steep temperature rise near $z=2$~Mm in panel (a) represents the transition region between the chromosphere and the corona.
We notice a temperature plateau of $T_{\rm gas}\approx 5000$~K appearing in the region $z=0.5 - 2$~Mm.
This temperature plateau is created as a result of the latent heat associated with \ion{H}{0}ionization, which acts as a thermostat, in particular for the shock waves propagating into the chromosphere \citep{carlsson+stein:1997,judge+:2010} when treated in statistical equilibrium \citep{carlsson+stein:2002} as done here.
%

%%% Describing time-averaged flux tube atmosphere in my simulation: Magnetic field
Figure\,\ref{fig:fluxtube-profile}b shows that the magnetic field becomes stronger than $1$~kG below the photosphere.
The magnetic field strength fluctuates widely at this depth as a result of the time variability of the flux-slab width---a clear characteristic of magnetic pumping.
Near the top boundary in the corona, on the other hand, the magnetic field strength stays at $45$~G with no significant fluctuation.  
%

%%% Describing time-averaged flux tube atmosphere in my simulation: density and pressure
The density and the gas pressure in the atmosphere of the flux slab are plotted in the panels (d) and (e), respectively.
We notice a plateau of enhanced density at the same height as of the temperature plateau, indicating the thermostatic behavior of propagating shock waves.
On the other hand, it is difficult to identify this same effect in both the magnetic field strength and the gas pressure in the panels (b) and (e), respectively.
The pressure scale height is plotted as a function of height in panel (f).
The chromospheric temperature plateau, visible in panel (a), creates a constant pressure scale height of about $130$~km in the chromosphere.
%

%%% Describing time-averaged flux tube atmosphere in my simulation: Cut-off frequency
The local cut-off frequency for longitudinal waves of the flux-slab atmosphere, $\nu_{\rm v}=\omega_{\rm v}/2\pi$, and that for transverse kink waves, $\nu_{\rm k}=\omega_{\rm k}/2\pi$, are plotted as a function of height in panel (c) in red and blue colours, respectively.
The cut-off frequency for longitudinal waves in the photosphere above $z=0$ is larger than $5\,{\rm mHz}$ and therefore larger than the conventional acoustic cut-off frequency $\nu_{\rm a}=3.3$~mHz of the magnetic field-free photosphere.
This discrepancy is a consequence of the strong magnetic field in the flux slab, within which the height of optical depth unity $\tau_{\rm 500}=1$ is systematically shifted downward by approximately $\Delta z=200$~km compared to the non-magnetized atmosphere, as a result of the Wilson effect.
This causes the cut-off frequency of the flux-slab atmosphere at photospheric heights of the non-magnetized atmosphere to be close to the %nominal
chromospheric acoustic cut-off-frequency of %the latter atmosphere of 
$\nu_{\rm a}=5.5$~mHz.
The cut-off frequency has a plateau around $5$~mHz in the chromosphere, which comes again as a result of the thermostat effect.
We note that the cut-off frequencies for longitudinal waves and also for transverse waves in the flux slab, as evaluated according to Equations\,(\ref{eqn:cutoff_longitudinal}) and (\ref{eqn:cutoff_transverse}) respectively, could differ from the actual cut-off frequency of a flux tube/slab structure with funnel-like expansion with height as a result of the ramp effect \citep[][see also Figure\,\ref{fig:snapshot}]{cally:2007}, but also because Equations\,(\ref{eqn:cutoff_longitudinal}) and (\ref{eqn:cutoff_transverse}) presume an isothermal hydrostatic atmosphere, small amplitude waves, and the thin tube approximation.

%%% Figure 5: Space-time diagram of the rate of diameter change
\begin{figure*}
\epsscale{1.0}
\plotone{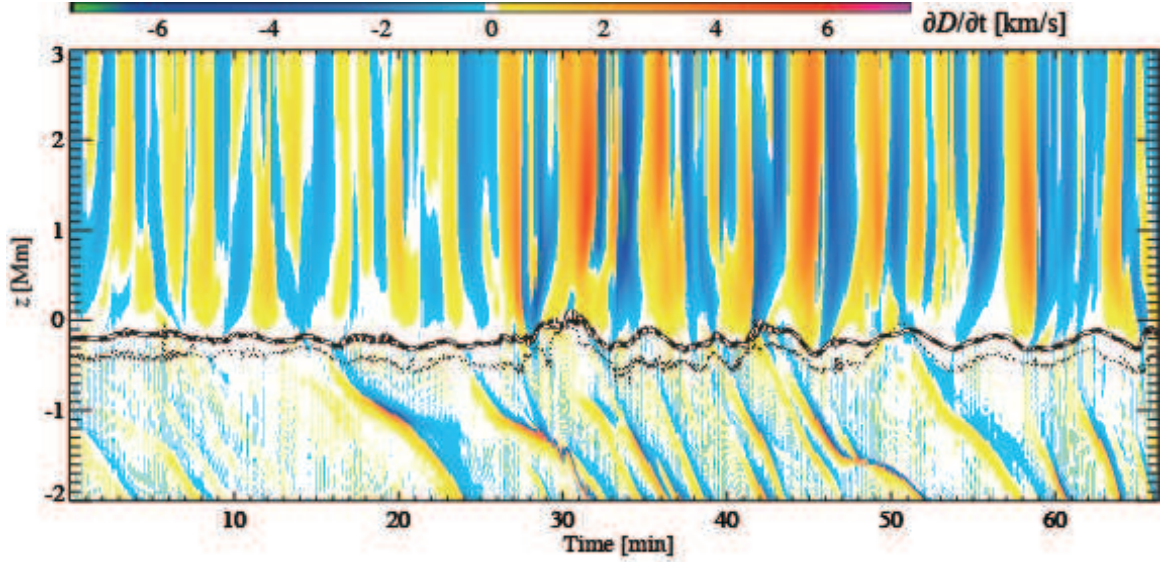}
\caption{Space-time diagram of the change of width of the flux slab.  The dashed black and white curve corresponds to the $\tau_{\rm 500}=1$ surface in the core region of the flux slab.  The dotted black curve corresponds to the $\beta=1$ surface in the core region of the flux slab.}
\label{fig:diameter}
\end{figure*}

\subsection{Onset of the magnetic pumping}
\label{sec:onset}

%%% Explain magnetic flux tube structure
Figure\,\ref{fig:diameter} shows the time-space diagram of the rate of width change of the flux slab, $\partial D/\partial t$.
As soon as the simulation starts, a weak constriction (blue) of the flux slab, associated with a convective downdraft (a weak magnetic pumping event) occurs in the lower part of the simulated convection layer, drifting downwards.
It is preceded by an expansion (yellow) of the flux slab caused by the same downwardly drifting convection cell.
A strong magnetic pumping event starts near the surface of optical depth unity (the black and white dashed curve) at time $t=16$ minutes, which can be followed for about $8$ minutes.
For a more detailed description of a pumping event, see Figure 2 of \citet{kato+:2011}.
Later on, similar events occur sporadically at roughly every 10 minutes.
The amplitude of the width-change rate in the convection zone is less than $\pm 10~{\rm km\,s^{-1}}$ and the downward drift speed of each convective cell is not more than $7~{\rm km\,s^{-1}}$ (while the actual plasma speed is usually supersonic). 
These pumping processes are responsible for the excitation of disturbances above the optical surface in the photosphere, and these disturbances propagate through the chromosphere to the corona.
We note that the fastest expansion and constriction, $\partial D/\partial t\geq\pm 6~{\rm km\,s^{-1}}$, occurs at $t=32$ minutes in the chromosphere ($z=1$~Mm).
This event seems to be associated with a transient transverse wave rather than with a pumping event and it happens only once during the entire time period of the simulation.

\subsection{Responses of the flux-sheath atmosphere}
\label{sec:response}
%%% Longitudinal and transverse wave propagations

%%% Figure 6: Space-time diagrams of the velocity
\begin{figure*}
\epsscale{0.75}
\plotone{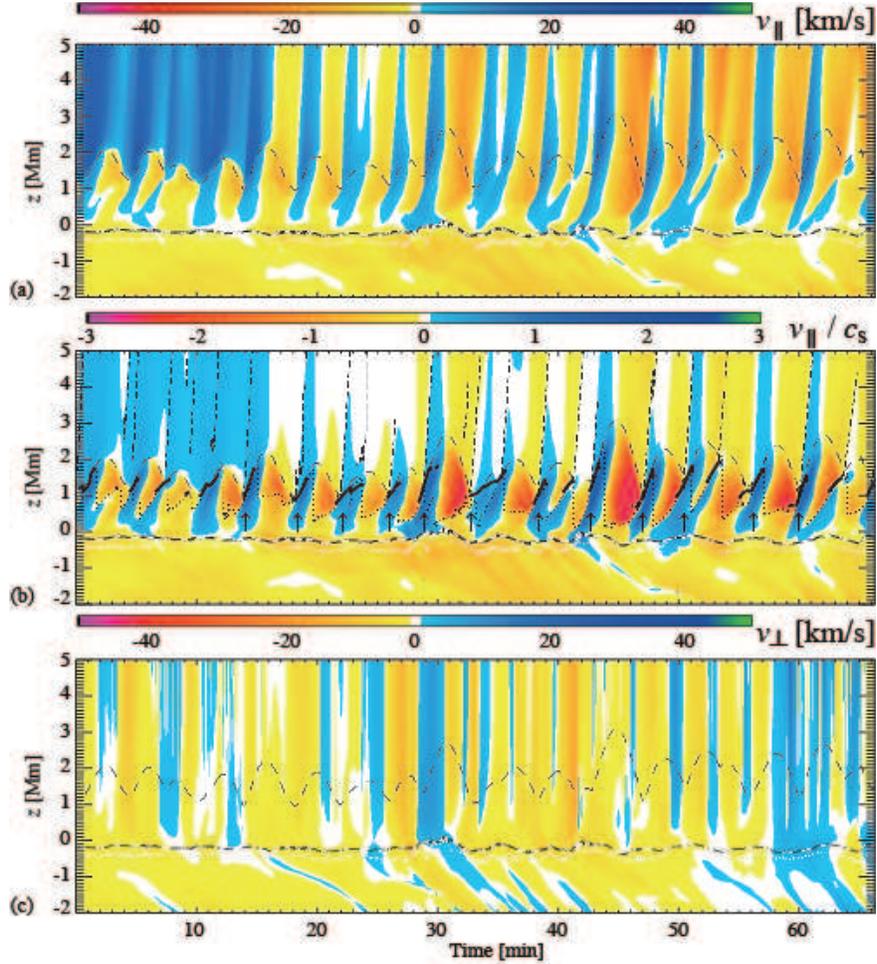}
\caption{Space-time diagrams of the velocity in the core of the flux slab.  Colours refer to (a) the velocity and (b) the Mach number parallel to magnetic field lines (longitudinal velocity) and (c) the velocity perpendicular to magnetic field lines (transverse velocity).  The dashed gray curve in all panels indicates the height of the transition region defined as $T_{\rm gas} = 100,000$ K.  The dashed black and white curve corresponds to the $\tau_{\rm 500}=1$ surface whereas the dotted white curve corresponds to the plasma-$\beta=1$ surface in the core of the flux slab.  Positive velocities indicate plasma motions in the upward and in the rightward direction for $v_{\parallel}$ and $v_{\perp}$, respectively.
In panel (b), black dotted curves and black dashed curves indicate the trajectories of the compressive waves in the lower atmosphere where $T_{\rm gas} < 5\times 10^{4}~{\rm K}$ and in the upper atmosphere where $T_{\rm gas} > 5\times 10^{4}~{\rm K}$, respectively.
Black solid curves indicate the trajectories of the strong shocks.}
\label{fig:propagation}
\end{figure*}

%%% Describe the longitudinal and transverse velocities in the flux tube
Figure\,\ref{fig:propagation} shows the space-time diagram of the velocity in the core of the flux slab for the full time period of 68 minutes of the simulation; on panels (a) and (b) the velocity parallel to the magnetic field, $v_{\parallel}$, and on panel (c) the velocity perpendicular to the magnetic field, $v_{\perp}$.
Panel (a) of Figure\,\ref{fig:propagation} reflects the upper atmospheric response to the repetitive magnetic pumping events taking place immediately beneath the surface of optical depth unity.
The upper atmosphere must respond to the magnetic pumping simply because of the frozen-in condition by which the atmosphere of the flux slab is governed.
Downflows inside the flux slab, driven by the magnetic pumping, cannot be replenished from the lateral direction but the plasma is forced to flow along the magnetic field from the upper layers.
As a result, a fast downflow occurs in the lower photosphere and a rarefaction propagates upward all the way through the chromosphere to the lower corona.
As soon as this downflow rebounds and compresses the gas inside the flux slab (because the pumping ceases or is being interrupted), it generates a slow-mode wave which also propagates upward and develops into a shock in the chromosphere.
Such a shock creates an arc-shaped temperature enhancement as visible in Figure\,\ref{fig:snapshot}\,b.
These shocks release the thermal energy that maintains the temperature plateau manifest in Figure\,\ref{fig:fluxtube-profile}\,a.
In panel (b), the trajectories of compressive body waves and shock waves are plotted over the time-space diagram of the local Mach number parallel to the magnetic field lines in the core of the flux slab.  In the chromosphere ($z\lsim 2$~Mm), most of the downward and upward velocities become supersonic.
In the photosphere, the velocity of downdrafts near the optical surface becomes almost supersonic several times, especially when strong magnetic pumping takes place such as at $t = 16$ minutes (see also Figure\,\ref{fig:diameter}).
We determine the location of compressive waves by tracing the location of minima of $\partial v_{\parallel}/\partial z$ as the precursors of shocks, and also determine the location of shocks by searching for pressure jumps where $\partial p_{\rm gas}/\partial z < -5\times 10^{-7}~{\rm [dyn\,cm^{-3}]} \sim - (p_{\rm gas,z=1\,Mm}/H_{\rm z=1\,Mm})$.
Note that there is always a velocity jump where a strong pressure jump exists.
The detected compressive waves (black dotted curve in panel (b)) 
are mostly associated with sonic points of downflows (rarefaction waves) and with sign changes of the velocity between downward in the upper part (rarefaction wave) and upward  in the lower part (slow-mode wave) of the flux slab. 
The detected shocks (black solid curve) extend the compressive waves and continue further into sonic points of upward velocity.
When the compressive wave reaches the transition region, it turns into a shock wave and occasionally creates an offspring in the form of a compressive wave as seen at $t\approx 18, 22, 26, 29, 33, 38, 43, 47, 50, 56, 60$, and $63$ minutes, indicated by arrows in panel (b).
The compressive waves propagate further into the corona, whereas the rest of the shocks dissipate still within the chromosphere/transition region ($z < 2$~Mm).

Panel (c) of Figure\,\ref{fig:propagation} shows mostly negative transverse velocities indicating that the flux slab drifts leftward. 
A sudden directional change of the drift motion appears just before $t = 30$ minutes all the way through the upper atmosphere above the surface of optical depth unity.
We note that the propagation of transverse waves is hardly visible in the chromosphere and corona from  panel (c) unlike the longitudinal wave in panel (a).
Below the surface of optical depth unity, on the other hand, transverse waves propagate downward similar to the longitudinal waves in panel (a).
Downwardly propagating waves in the convection zone are associated with a width change of the flux slab as shown in Figure\,\ref{fig:diameter}, and therefore with the convective speed of the downflow cells adjacent to the flux slab.
In Appendix\,\ref{sec:shocks}, we take a closer look at two selected time windows and follow the development of compressive waves into shocks arising from two distinctly different excitation mechanisms: longitudinal excitation in Figure\,\ref{fig:timing1} and transverse excitation in Figure\,\ref{fig:timing2}.

\subsection{Propagation speed of the compressive waves in the flux slab}
\label{sec:speed}

%%% Figure 10: propagation speed of shocks and waves
\begin{figure}
\epsscale{1.0}
\plotone{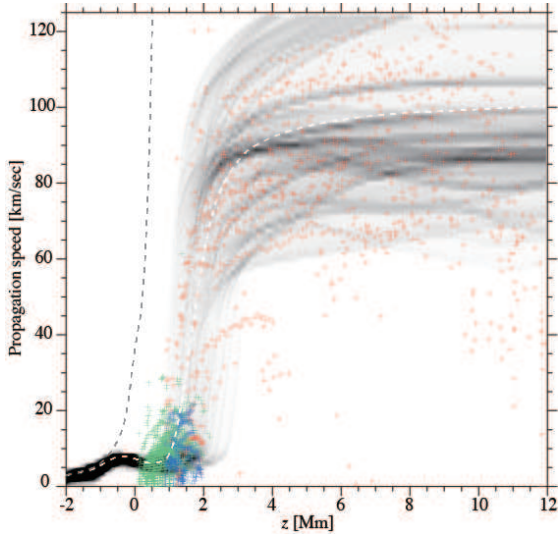}
\caption{Comparison between the longitudinal tube speed $c_{\rm T}$ and the propagation speed of shocks and compressive waves as a function of height.  The blue crosses indicate the propagation speed of strong shocks identified by a pressure jump (black solid curves in 
Figure\,\ref{fig:propagation}b) whereas the red and green crosses indicate the propagation speed of compressive waves identified by minima of $dv_{\parallel}/dz$ above and below $T=5\times 10^{4}$ K, respectively (black dotted and dashed curves in Figure\,\ref{fig:propagation}b).  The gray scales indicate the scatter of the longitudinal tube speed as computed with Equation\,(\ref{eqn:tubespeed}).  The white dashed curve indicates the time-averaged longitudinal tube speed whereas the grey dashed curve indicates the time-averaged Alfv\'en speed.}
\label{fig:fluxtube-ct}
\end{figure}

%%% Propagation speed of the compressive tube waves
Figure\,\ref{fig:fluxtube-ct} shows the distribution of the propagation speed of compressive waves and shocks, evaluated by taking the time derivative of the trajectories of the compressive waves and shocks of Figure\,\ref{fig:propagation}b.
By comparison, the distribution of the tube speed, $c_{\rm T}$, according to Equation~(\ref{eqn:tubespeed}), is shown in gray scales in the background.
Red and green marks indicate the propagation speed of compressive waves above and below $T=5\times 10^{4}$ K respectively, whereas blue marks indicate the propagation speed of shock waves.
The propagation speed is approximately consistent with the tube speed at all heights above the surface of optical depth unity.
This confirms that our detection of waves is reasonably good and also that the tube speed represents the propagation speed of compressive and shock waves in the photospheric and chromospheric layers of the flux slab reasonably well.
Note that the actual propagation speed could be slightly faster because the flux slab extends not always exactly in the vertical direction in the simulation.
Essentially, all these waves are predominantly acoustic in nature.

\subsection{Velocity amplitudes of the waves in the flux slab}
\label{sec:amplitude}

%%% Figure 11: Scatter of longitudinal and transverse velocity
\begin{figure*}
\epsscale{1.0}
\plottwo{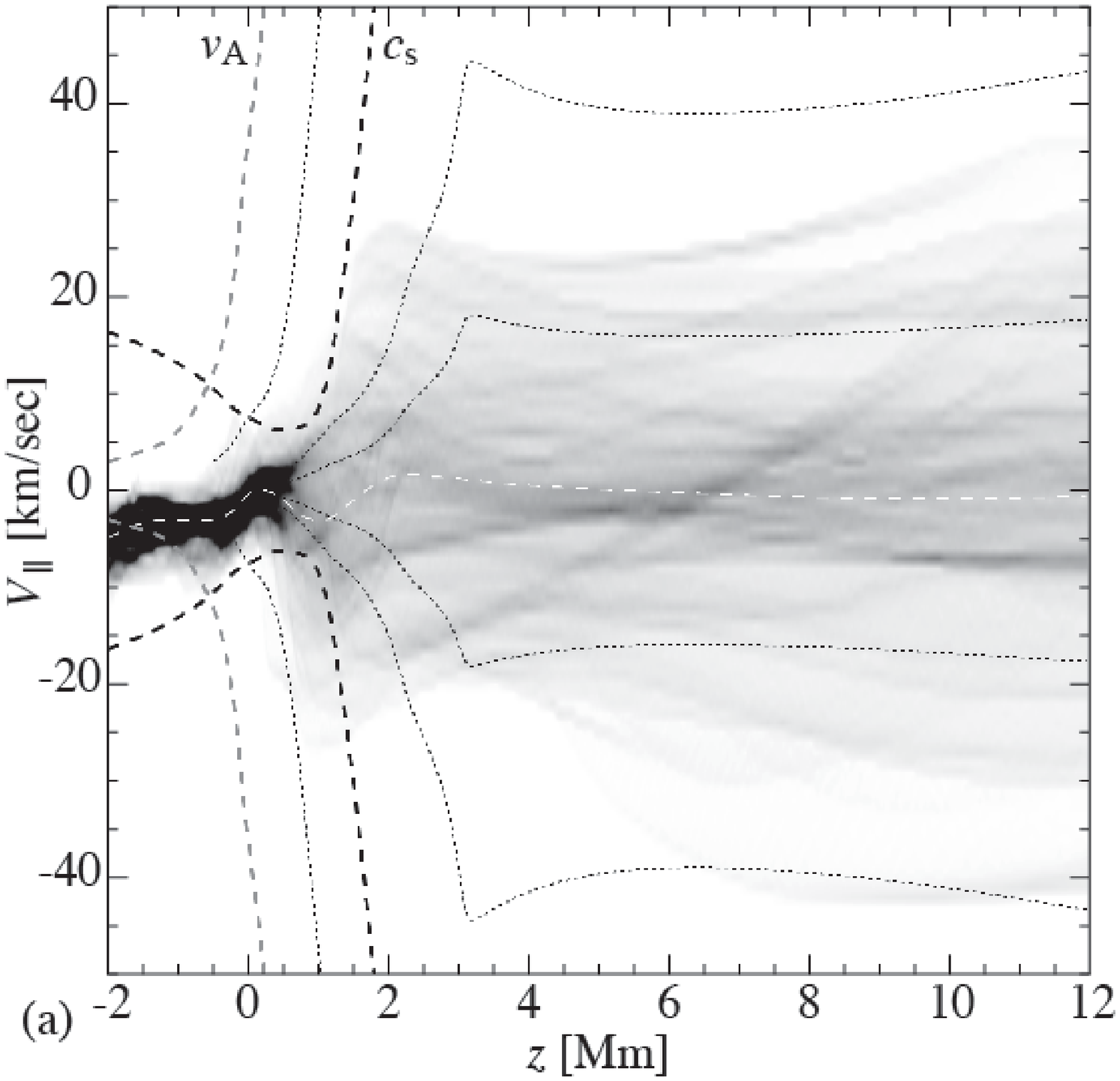}{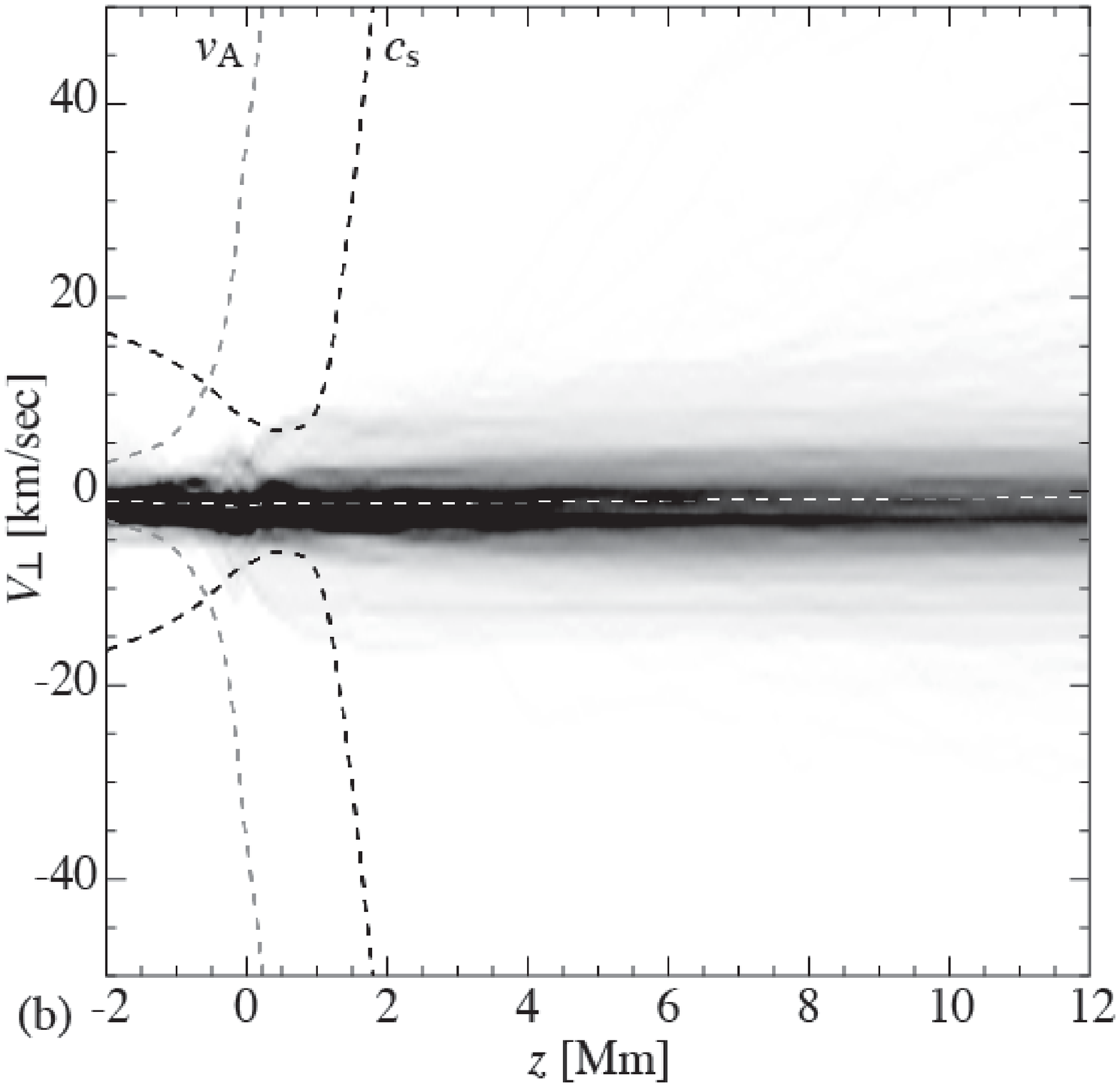}
\caption{Scatter plots of (a) the longitudinal velocity $v_{\parallel}$ and (b) the transverse velocity $v_{\perp}$ inside the flux slab in gray scales for the full time period of 68 minutes.  Time-averaged velocities, the local sound speed, and the local Alfv\'en speed are shown by white, black, and grey dashed curves respectively.  The three pairs of black dotted curves in panel (a) show the theoretical curves that result from conservation of longitudinal wave-energy flux inside the flux slab (Equation\, (\ref{eqn:wave_conservation})) starting from the mean velocity amplitude at three different reference heights near the optical surface, i.e. at $z=-0.5$, $-0.08$, and $0.0$~Mm.}
\label{fig:velocity_amplitude}
\end{figure*}

%%% Amplitude of longitudinal velocity
Figure\,\ref{fig:velocity_amplitude} shows the scatter in gray scales of (a) the longitudinal velocity and (b) the transverse velocity in the core of the flux slab over the full time period of 68 minutes.
It is obvious that the longitudinal waves dominate in the flux slab because the maximal longitudinal velocity amplitude is approximately $\pm 50~{\rm km\,s^{-1}}$ (the maximum root-mean-squared (rms) amplitude is $15~{\rm km\,s^{-1}}$) whereas the maximal transverse velocity amplitude is only approximately $\pm 15~{\rm km\,s^{-1}}$ (the maximum rms amplitude is $8.5~{\rm km\,s^{-1}}$).
If the maximum longitudinal velocity amplitude is due to longitudinal waves, it should be limited by the conservation of wave-energy flux in the flux slab:
\begin{equation}
\rho v_{\parallel}^{2}c_{\rm T} D = {\rm const.},
\label{eqn:wave_conservation}
\end{equation}
where $\rho$ and $D$ are the density and the width of the flux slab, respectively.
The tube speed $c_{\rm T}$ is computed from Equation\,(\ref{eqn:tubespeed}) using the time-averaged flux-slab atmosphere of Figure\,\ref{fig:fluxtube-profile}.
The three pairs of black dotted curves in panel (a) show the theoretical values for the conservation of longitudinal wave-energy flux inside the flux slab starting from the mean velocity amplitudes at three different reference heights near $z=0$, i.e., $z=-0.5$~Mm, $z=-0.08$~Mm, and $z=0.0$~Mm.
The surface of optical depth unity inside the flux slab is roughly at $z=-0.2$~Mm due to the Wilson depression whereas the plasma-$\beta=1$ surface is deeper than $z=-0.5$~Mm.
From panel (a) we see that the maximal negative velocity amplitudes in the height range from $z=-0.5$~Mm to $z=1.0$~Mm are consistent with the theoretical curve for the conservation of wave-energy flux starting at $z=-0.5$~Mm.
They are due to rarefaction waves starting below the surface of optical depth unity, initiated by magnetic pumping (see schematic of Figure\,\ref{fig:schematic}).
They become larger than the local sound speed, which is indicated by the black dashed curves.
Such supersonic downflows lead to strong compression generating upwardly propagating slow modes in the flux slab.
The maximal positive velocity amplitudes up to $z=2$~Mm, on the other hand, are regulated by the local sound speed.
They are due to upwardly propagating slow waves driven by the rebound of the preceding downflows of the rarefaction.
Above $z=2$~Mm, the longitudinal velocity amplitudes in both upward and downward direction tend to be limited by the conservation of wave-energy flux starting at the reference height immediately below the surface of optical depth unity at $z=-0.08$~Mm.
This indicates that the waves of maximal amplitude seen in the upper chromosphere originate from that height, where the rebound takes place.
%

%%% Amplitude of transverse velocity
In panel (b) of Figure\,\ref{fig:velocity_amplitude}, the time-averaged transverse velocity amplitude (the white dashed curve) is slightly negative at all heights because of the predominantly leftward drift of the flux slab.
The transverse velocity becomes supersonic near the temperature minimum at $z\approx 1$~Mm but it never becomes super-Alfv\'{e}nic (except for the deep layer in the upper convection zone where $\beta\gg 1$).
Since the transverse velocities are significantly smaller than the longitudinal ones, the transverse waves are not the dominant carrier of wave-energy flux in the flux slab.
Mode conversion from transverse waves to longitudinal waves is possible \citep[e.g.,][]{ulmschneider+:1991,hasan+:2003}, but there is no obvious signature of it in the panels (a) and (b).
Unlike the longitudinal velocity amplitude in panel (a), the maximal amplitude of the transverse velocity is not significantly increasing above the surface of optical depth unity.
This indicates that the transverse velocity more likely originates in the upper photosphere or in the chromosphere directly rather than in the boundary between the upper convection zone and the photosphere (see also Figure\,\ref{fig:timing2} and corresponding text in Appendix\,\ref{sec:shocks}).

\subsection{Power spectra of the waves in the flux slab}
\label{sec:power}

%%% Figure 12: Wavelet power
\begin{figure*}
\epsscale{1.0}
\plotone{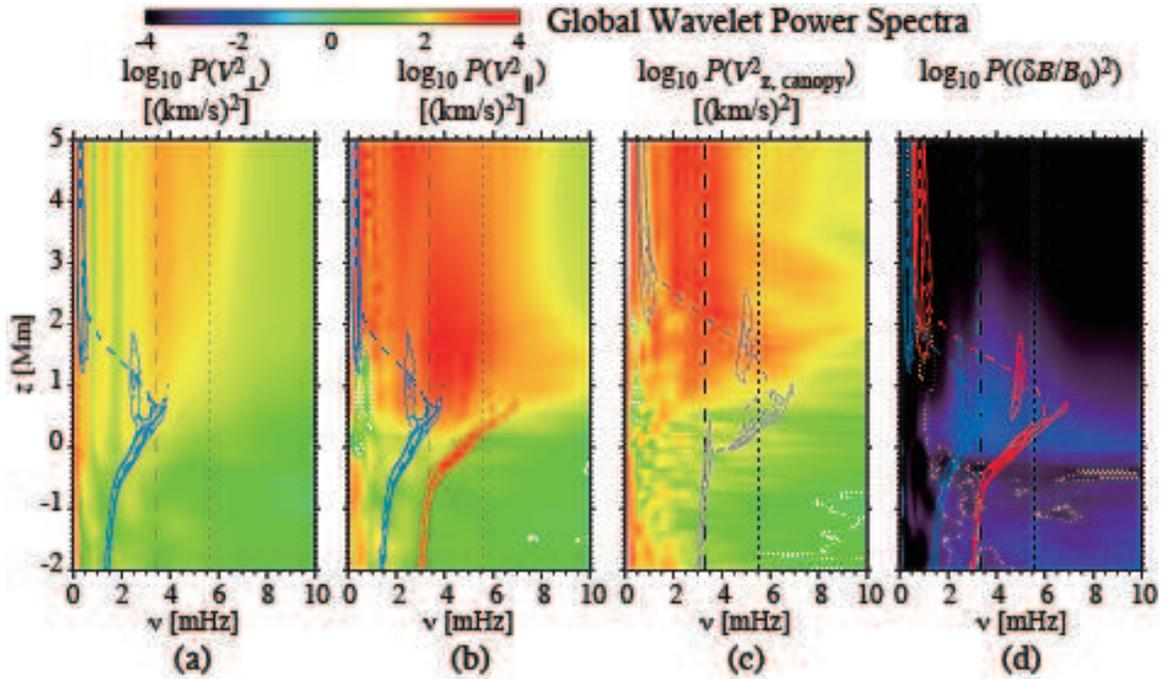}
\caption{The wavelet power spectra averaged over the full time period of 68 minutes.  The colour-scale refers to the power of (a) the transverse velocity, $v_{\perp}$, (b) the longitudinal velocity, $v_{\parallel}$, both in the core of the flux slab, (c) the vertical velocity outside the flux-slab atmosphere, $v_{\rm z, \rm canopy}$, for comparison, and (d) the magnetic flux density fluctuation $\delta B(t)\equiv B(t+\delta t) - B(t)$, again in the core of the flux slab, normalised by the time-averaged magnetic flux density, $B_{\rm 0}$, all plotted using logarithmic scales.  The white dotted contour encloses regions of smaller than $95\%$ confidence where the power is relatively small.  The blue contours indicate the local cut-off frequency distribution for the transverse kink waves $\omega_{\rm k}/2\pi$ while the blue dashed curve shows its time-averaged profile.  The red contours indicate the cut-off frequency distribution of the longitudinal wave $\omega_{\rm v}/2\pi$ while the red dashed curve shows its time-averaged profile.  The gray contours indicate the acoustic cut-off frequency, $\omega_{\rm a}/2\pi=c_{\rm s}/2H$.  The black vertical dashed and dotted lines correspond to 5 minute and 3 minute periods, respectively.}
\label{fig:global_wavelets}
\end{figure*}

%%% Longitudinal waves dominate the flux tube
The global wavelet power spectra of the flux-slab atmosphere for the full time period are shown in Figure\,\ref{fig:global_wavelets}.
The power spectra of the longitudinal and transverse velocities and of the magnetic field-strength fluctuations are calculated by using the Morlet wavelet analysis technique of \citet{torrence+compo:1998}.
The local cut-off frequency of  the longitudinal waves and that of transverse kink waves, evaluated according to Equations\,(\ref{eqn:cutoff_longitudinal}) and (\ref{eqn:cutoff_transverse}), are shown by red and blue contours, respectively (see also Figure\,\ref{fig:fluxtube-profile}c).
By comparing the time-averaged wavelet power spectra of the transverse velocity in panel (a) and that of the longitudinal velocity in panel (b), both spatially averaged over the core of the flux slab, it is obvious that the longitudinal waves dominate in the flux slab.
In panel (a),  the transverse waves are prominent near $3.3$~mHz (5 minutes period marked with a dashed black line) in the upper part of the atmosphere, which also corresponds to the maximum cut-off frequency for transverse kink waves in the photosphere.
%

%%% But longitudinal waves are puzzling
In contrast to panel (a), panel (b) shows that the longitudinal wave power occurs in a wide frequency band between $1$ - $10$~mHz at $z\gsim 0.3$~Mm.
The power of the oscillations concentrates at $\nu\approx 4$~mHz between $z=1$ and $3$~Mm, which is distinctive from the power of vertical oscillations $4$~Mm sideways of the flux slab in panel (c).
The maximum power of longitudinal waves is $3.3$ in logarithmic scale, corresponding to a velocity amplitude of $49~{\rm km\,s^{-1}}$, which is consistent with the maximal longitudinal velocity amplitudes shown in Figure\,\ref{fig:velocity_amplitude}a.
It is striking that the frequency of maximum power of the longitudinal waves is to a large degree below the local cut-off frequency of longitudinal waves.
This is probably due in part to the ramp effect \citep{cally:2007} by which longitudinal slow modes below the cut-off frequency can still propagate where the magnetic field is inclined with respect to the vertical direction \citep{michalitsanos:1973,bel+leroy:1977,jefferies+:2006,de_pontieu+:2004,heggland+:2011}.
On the other hand, one should also keep in mind that Equation\,(\ref{eqn:cutoff_longitudinal}) for computing the local cut-off frequency breaks down for the highly non-linear waves that we deal with here.
%

%%% The ramp effet
The ramp effect definitively plays an important role for the longitudinal velocity power shown in panel (c).
This power was evaluated for a vertical line-of-sight, $4$~Mm sideways of the flux slab but still within the reach of its magnetic canopy, which starts there at a height of $z\sim 1$~Mm.
Above this height, $\beta\ll 1$ (see also Figure\,\ref{fig:snapshot}) so that the slow modes detected there are guided by the magnetic field and have propagated along the strongly inclined magnetic fields of the peripheral, expanding region of the flux slab below $z\sim 1$~Mm.
The characteristic frequency is strongly modified by the inclination angle of magnetic fields in chromospheric and coronal heights at this location and therefore the power in panel (c) concentrates around $3.3$~mHz.
This behaviour can also be seen in Figure\,14 of \citet{heggland+:2011}.
%

%%% magnetic oscillation in the photosphere
Panel (d) shows the power of the magnetic oscillations.
The overall amplitude is roughly $-2$ in logarithmic scale, that is $\delta B(t)/\langle B\rangle\approx 0.1$ or $\delta P_{\rm mag}(t)/\langle P_{\rm mag}\rangle\approx 0.01$.
The power is concentrated at $4$~mHz which is identical to the peak frequency of the longitudinal waves in the panel (b) and also extends from the surface of optical depth unity, $z=0$~Mm, to the transition region $z\approx 3$~Mm, indicating a connection between magnetic oscillations and longitudinal waves.
There is also significant power in a wide frequency range from $1$ to $10$~mHz in the photospheric layers.

%%% Local wavelet power spectra
In Appendix\,\ref{sec:local_power}, the time-dependent wavelet power spectra are shown as a function of time at different heights.
It confirms the transient nature of the transverse oscillations and the persistent, resonant nature of the longitudinal oscillations (see Figure\,\ref{fig:local_wavelets}).

\section{Implications, shortcomings, and discussion of the results}
\label{sect4}

This section discusses the low power of transverse oscillations in the present simulation, energy dissipation rates, and various shortcomings of the simulation.

\subsection{Longitudinal waves vs. transverse waves}
A remarkable finding is that longitudinal waves dominate the flux-slab atmosphere, rather than transverse waves as indicated by the global and local wavelet power spectra (see Figures\,\ref{fig:global_wavelets} and  \ref{fig:local_wavelets}, respectively).
One may ask why transverse waves are less powerful and less frequent, even though it was shown in previous studies that rapid foot point motions of flux tubes are able to excite transverse waves of considerable power \citep[e.g.,][]{choudhuri+:1993a, hasan+kalkofen:1999, cranmer+van_ballegooijen:2005}.
Our numerical simulation shows that the transverse motion of the flux slab in the upper convection zone is not rapid enough to generate high frequency transverse waves ($\nu > 2$~mHz) and it occurs not frequently (only once during the full time period of $68$ minutes).
This result is discussed further below.
%

%%% consistency with Bogdan et al. 2003
The present result is consistent with the results of \citet{bogdan+:2003} or \citet{vigeesh+:2009} who investigated the generation of magneto-acoustic waves depending on the driving force at the foot-point of the flux slab and depending on the magnetic field strength.
In the present simulation, however, the longitudinal and transverse driving is not introduced by ad hoc piston and motions but by the action of self-consistent magnetoconvective motion.
According to the terminology of \citet{bogdan+:2003}, the present simulation corresponds to the case of ``strong magnetic field with radial driving''  because direct longitudinal excitation via magnetic pumping occurs more frequently than transverse disturbance.
%

%%% a problem of lack of transverse waves requires new excitation mechanism
The lack of transverse waves raises a serious problem because recent observations suggest that there are plenty of large velocity amplitude transverse waves in the solar atmosphere %(see Table\,1).
\citep{fujimura+tsuneta:2009,kuridze+:2012,stangalini+:2013a,morton+:2014,stangalini+:2015}.
Obviously, some driving mechanisms for generating persistent transverse waves in the real Sun must exist.
A hint may already be given in Figure\,\ref{fig:timing2} where transverse waves are seen to be generated in our simulation.
We expect that multiple flux concentrations within the same computational domain would more frequently generate transverse waves than a single flux slab.
It may also be possible that transverse waves are generated more efficiently in regions with weaker fields or other field topologies.
Last but not least, we must be aware that the limitation to two spatial dimensions also reduces the degree of freedom for transverse waves and excludes shear Alfv\'en waves completely.
It is yet to be confirmed what is the dominant process for generating waves in three spatial dimensions.
When the flux tube is located in a velocity shear, a swirling motion may evolve and generate a 'magnetic tornado' \citep{wedemeyer-bohm+:2012} and/or propagating twist \citep{de_pontieu+:2014b}.
Such events are  expected to occur sporadically, perhaps similar to the generation of large amplitude transverse waves in our study.
On the other hand it is exactly this limitation to two spatial dimensions that provides us the advantage of obtaining a stable flux sheet over a sufficiently long time period for computing power spectra like those of Figures\,\ref{fig:global_wavelets} and \ref{fig:local_wavelets}.
%

%% Table removed from the LaTeX file !!

%%% acoustic mode leakage
Based on the assumption that only acoustic waves above the cut-off frequency can propagate from the upper convection zone to the upper atmosphere, the transverse waves should have more power there than the longitudinal waves because the cut-off frequency of the transverse waves is smaller than that of the longitudinal waves \citep{spruit:1981}.
However, leakage of acoustic modes into the photosphere and chromosphere below the nominal cut-off is possible along inclined magnetic field lines as was confirmed by different groups \citep[e.g.][]{hansteen+:2006, jefferies+:2006, heggland+:2011}.
Our result indicates that in magnetic flux concentrations of predominant vertical orientation such as found in magnetic bright points (MBPs) and in the network regions, magnetic pumping provides a viable source of longitudinal compressive waves at chromospheric and coronal heights, additional to the leakage of acoustic modes of the global oscillations.

\subsection{Energy dissipation and energy fluxes}

%%% Figure 14: Space-time of energies
\begin{figure*}
\epsscale{0.9}
\plotone{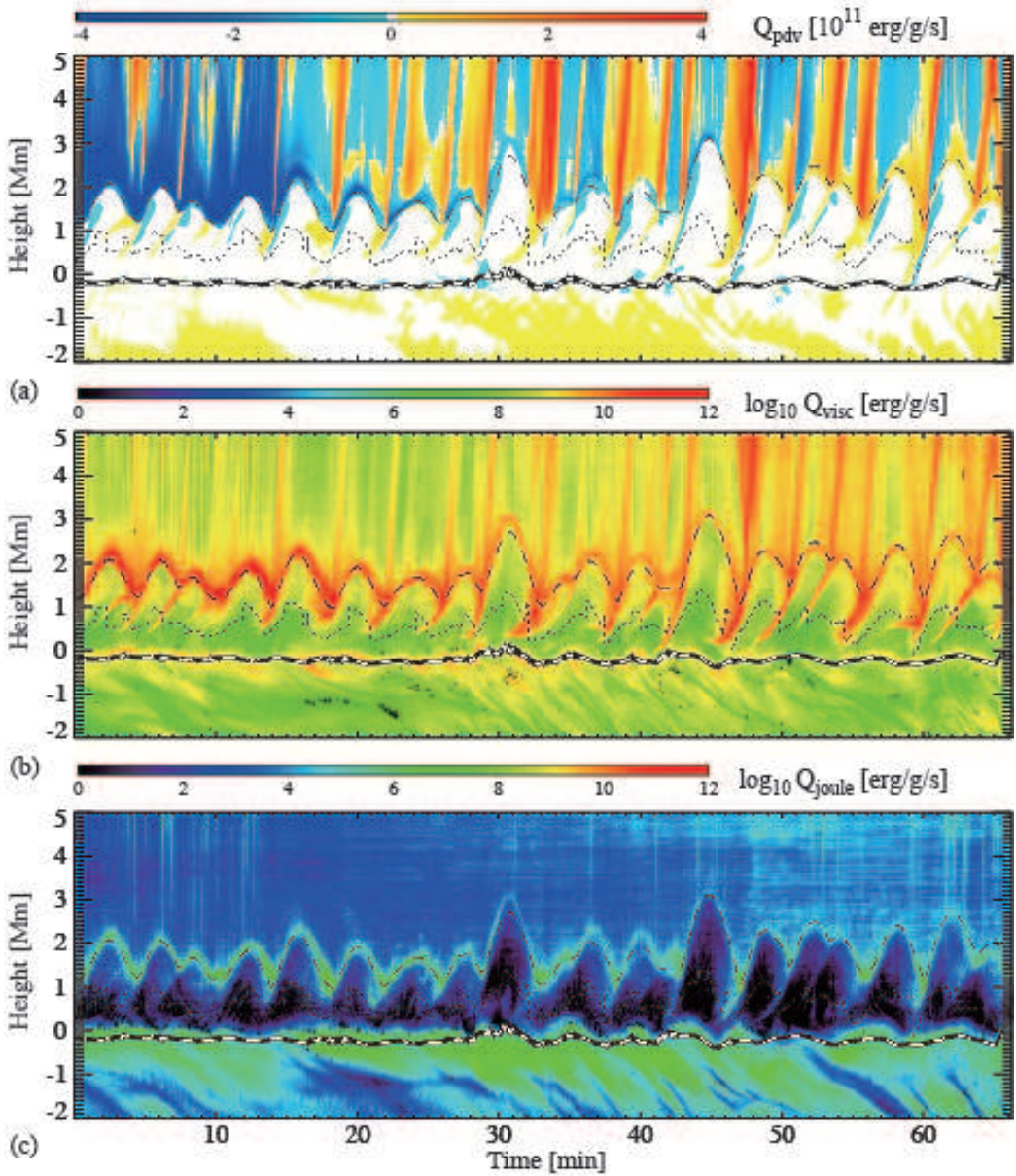}
\caption{Time-space diagram of (a) the heating/cooling rate due to compression and expansion of the gas, (b) the energy dissipation rate by viscous heating, and (c) the energy dissipation rate by Joule heating.  The dashed black and white curve corresponds to the $\tau_{\rm 500}=1$ surface.  The dotted and dashed grey curves corresponds to the height of the instantaneous temperature minimum and the height of $T_{\rm gas}=100\,000$ K in the core of the flux slab, respectively.  Note that all of these quantities are horizontally averaged inside the core of the flux slab.}
\label{fig:energetics}
\end{figure*}

%%% energy deposition
Figure\,\ref{fig:energetics} shows the time evolution of the energy deposition rates per gram of (a) the work done by compression and expansion $Q_{\rm pdv}$, (b) viscous heating $Q_{\rm visc}$, and (c) Joule heating $Q_{\rm Joule}$ in the core of the flux slab.
Quasi-adiabatic compression and expansion occurs mainly in the lower corona  ($z\geq 2$~Mm) in an intermittent fashion associated with the compressive and rarefaction waves.
The heating/cooling is on the order of $Q_{\rm pdv}\sim 10^{11}~{\rm erg\,g^{-1}\,s^{-1}}$ as shown in panel (a).
The viscous heating in panel (b) occurs mainly in the transition region between chromosphere and corona due to shock dissipation, but it also becomes prominent along the trajectories of compressive waves in the corona (see also Figure\,\ref{fig:propagation}b). %Figure\,\ref{fig:shocks}). 
The viscous heating rate is on the order of $Q_{\rm visc}\sim 10^{10}~{\rm erg\,g^{-1}\,s^{-1}}$.
Viscous heating also occurs near the surface of optical depth unity but its value there is one order of magnitude lower than that in the transition region.
In panel (c), Joule heating occurs both in the upper convection zone just below the surface of optical depth unity ($z\leq 0$~Mm) and also in the transition region.
The Joule heating rate is on the order of $Q_{\rm Joule}\sim 10^{8}~{\rm erg\,g^{-1}\,s^{-1}}$, which is much less than that of the viscous heating rate. 
%

%%%
While the spatial locations where major dissipations occur in Figure\,\ref{fig:energetics} are plausible, the sharpness of their confinement, amplitude, and relative strength depend on the inherent numerical dissipation of the simulation, which strongly varies over the computational domain because no realistic dissipation is in use but rather a scheme for achieving stability with least viscosity and resistivity.
Therefore, energy fluxes resulting from the spatiotemporal integration of the energy-dissipation rates $Q$ may be more meaningful. Thus,
\begin{equation}
F=\frac{1}{{\cal L}{\cal T}}\int\!\! {\rm d}t\int\!\! {\rm dz}
 \int \!\! {\rm d}x\; Q(x,z,t)\,\rho(x,z,t)\,,
\label{eqn:energyflux}
\end{equation}
where ${\cal L}=11.2$~Mm and ${\cal T}=68$ minutes are the horizontal simulation box size and the total simulation time, respectively.
We integrate Equation\,(\ref{eqn:energyflux}) in the horizontal direction over the core of the flux slab of width $D(z,t)$ only.
In the vertical direction, we integrate from the surface of optical depth unity to the height of the instantaneous temperature minimum for the photosphere, from the height of the instantaneous temperature minimum to the height of $T_{\rm gas}=100\,000$ K for the chromosphere, and above the height of $T_{\rm gas}=100\,000$ K for the corona.
For the integration over the chromosphere, we obey the additional restriction that the temperature must surpass the chromospheric temperature plateau of $T_{\rm gas}=5000$ K.
The energy deposition above the height of the instantaneous temperature minimum but below $T_{\rm gas}=5000$ K is then counted as photospheric.
The determined energy fluxes in each of the three atmospheric layers are summarized in Table\,1.
In the chromosphere, the energy flux resulting from viscous heating alone of $1.6\times 10^{4}~{\rm W\,m^{-2}}$ is sufficient to compensate for the empirical radiative loss in the chromosphere of $2500 - 3300~{\rm W\,m^{-2}}$ \citep{ulmschneider:1974} or $4300~{\rm W\,m^{-2}}$ \citep{vernazza+:1981}, or for the estimated energy flux of magneto-acoustic waves in the upper photosphere of $6400 - 7700 {\rm W\,m^{-2}}$ by \citet{bello_gonzalez+:2010}.
Since we integrate in Equation\,(\ref{eqn:energyflux}) only over the core of the flux slab but divide instead by the full computational box width, the numbers in Table\,1 refer to an arbitrary spatial filling factor, given by the simulation setup.
Here the flux-slab has a width of $100$ km at the base of the photosphere and $1$~Mm in the chromosphere, and therefore the photospheric filling factor is $\sim 0.01$ while the chromospheric filling factor is $\sim 0.1$.
For other filling factors, the fluxes in Table\,1 need to be scaled correspondingly. 
Clearly, viscous dissipation due to compressive and shock waves is the most important contribution to the dissipative energy fluxes in the chromosphere and also in the corona.
However, this result may change in favor of Joule dissipation in environments of more complex, three-dimensional magnetic field structures.
%

%% The values (usually only l,r and c) in the last part of
%% \begin{deluxetable}{} command tell LaTeX how many columns
%% there are and how to align them.
\begin{deluxetable}{rccccc}
%% Rotate to a landscape orientation
%\rotate
%
%% Over-ride the default font size
%% Use Default (12pt)
%
%% Use \tablewidth{?pt} to over-ride the default table width.
%% If you are unhappy with the default look at the end of the
%% *.log file to see what the default was set at before adjusting
%% this value.
\tabletypesize{\scriptsize}
%% This is the title of the table.
\tablecaption{The energy fluxes resulting from the various energy deposition rates at three solar atmospheric layers.  }
\label{table:energy_fluxes}
%
%% This command over-rides LaTeX's natural table count
%% and replaces it with this number.  LaTeX will increment 
%% all other tables after this table based on this number
\tablenum{1}
\tablewidth{0pt}
%
%% The \tablehead gives provides the column headers.  It
%% is currently set up so that the column labels are on the
%% top line and the units surrounded by ()s are in the 
%% bottom line.  You may add more header information by writing
%% another line between these lines. For each column that requries
%% extra information be sure to include a \colhead{text} command
%% and remember to end any extra lines with \\ and include the 
%% correct number of &s.

\tablehead{\colhead{} & \colhead{$F_{\rm vis}$} & \colhead{$F_{\rm Joule}$} & \colhead{$F_{\rm pdv}$} & \colhead{$F_{\rm vis}+F_{\rm Joule}+F_{\rm pdv}$} \\
\colhead{} & \colhead{(${\rm W\,m^{-2}}$)} & \colhead{(${\rm W\,m^{-2}}$)} & \colhead{(${\rm W\,m^{-2}}$)} & \colhead{(${\rm W\,m^{-2}}$)}}

%% All data must appear between the \startdata and \enddata commands
\startdata
Corona
& $7.4\times 10^{2}$
& $4.6\times 10$
& $-2.9\times 10^{3}$ 
& $-2.1\times 10^{3}$ \\
Chromosphere
& $1.6\times 10^{4}$
& $4.6\times 10^{2}$
& $-7.3\times 10^{3}$ 
& $9.0\times 10^{3}$ \\
Photosphere
& $6.3\times 10^{6}$
& $8.0\times 10^{6}$
& $-2.5\times 10^{6}$
& $1.2\times 10^{7}$ \\
\enddata

%% Include any \tablenotetext{key}{text}, \tablerefs{ref list},
%% or \tablecomments{text} between the \enddata and 
%% \end{deluxetable} commands

%% General table comment marker
%\tablecomments{Not yet.}

%% \tablerefs indicated
% \tablenotetext{a}{\citet{fujimura+tsuneta:2009}}
% \tablenotetext{b}{\citet{morton+:2012}}
% \tablenotetext{c}{\citet{stangalini+:2013a}}
% \tablenotetext{d}{\citet{morton+:2014}}
% \tablenotetext{e}{\citet{stangalini+:2015}}

\end{deluxetable}

\section{Conclusions}
\label{sect5}
%%%
In this paper we have investigated the generation and propagation of chromospheric and coronal waves in an isolated slab of magnetic flux, driven by magneto-convective processes in the deep photosphere and beneath it.
Using Bifrost simulations, we have confirmed the existence of  the magnetic pumping process, first seen in CO\raisebox{0.5ex}{\footnotesize 5}BOLD simulations by \citet{kato+:2011}.
Such pumping events generate slow modes that propagate upward and develop into shock waves in the chromosphere.
Once reached by the shock wave, the transition region is pushed up, which generates a compressive wave in the corona.
The disturbance caused by episodic magnetic pumping events is strong enough to maintain oscillations of the flux-slab atmosphere with a period of $\sim 4$ minutes at chromospheric and coronal heights during almost the full time period of the simulation of $68$ minutes.

%%%
By examining the properties of chromospheric/coronal oscillations in the flux slab, such as the propagation speed, the amplitude of both longitudinal and transverse velocities, and their characteristic frequencies, we find that the analytical solution of the Klein-Gordon type Equation~(\ref{eqn:Klein-Gordon_longitudinal}) is adequate for representing basic properties of longitudinal waves of the dynamic flux-slab atmosphere.
The propagation speed of longitudinal waves is thus regulated by the tube speed  $c_{\rm T}$ (Section\,\ref{sec:speed}).
The maximal longitudinal velocity amplitude is constrained by the conservation of wave-energy flux inside the flux slab from the deep photosphere to the corona (Section\,\ref{sec:amplitude}).
The velocity amplitude of the transverse waves is distinctively smaller than that of the longitudinal waves.
The characteristic frequency of transverse oscillations is consistent with the cut-off frequency of transverse kink waves in the chromosphere, but the transverse oscillations are transient, lasting for a period of less than $20$ minutes (Section\,\ref{sec:power} and Appendix\,\ref{sec:local_power}).
Longitudinal oscillations, on the other hand, are stable for more than $50$ minutes, having a characteristic frequency of $4$~mHz instead of the cut-off frequency of $3.3$~mHz of the corresponding static flux-slab atmosphere (Section\,\ref{sec:power}).
%

%%% Observational remarks
%
We conclude that magnetic pumping is a robust mechanism for generating chromospheric and coronal waves in the vicinity of strong magnetic flux concentrations.
While an observational detection of the magnetic pumping mechanism has not been reported yet, the simultaneous recording of vector magnetograms and Doppler velocities of suitable photospheric and chromospheric lines in the close surroundings of magnetic elements at the highest possible spatial/temporal resolution should make it possible.
It should allow us to detect variations in diameter and field strength of magnetic elements in association with the downflow jets in their immediate vicinities and subsequent responses in the chromosphere and the transition region.
Such observing programs would open new ways to the exploration of the source of dynamic fibrils and of chromospheric/coronal emissions.

\begin{acknowledgements}

This work was started and supported in part by the JSPS fund \#R53 (``Institutional Program for Young Researcher Overseas Visits'', FY2009-2011) allocated to NAOJ, part of which is managed by Hinode Science Center, NAOJ.  YK expresses sincere thanks to Saku Tsuneta, Toshifumi Shimizu, and Yoshinori Suematsu for providing strong support and encouragement.  YK and SW acknowledge support by the Research Council of Norway, grant 221767/F20.  The research leading to these results has received funding from the European Research Council under the European Union's Seventh Framework Programme (FP7/2007-2013) / ERC grant agreement No. 291058.  This research was supported by the Research Council of Norway through the grant ``Solar Atmospheric Modeling'' and through grants of computing time from the Programme for Supercomputing.  We thank the anonymous referee for many detailed comments on the manuscript of this paper.
\\
\end{acknowledgements}

%%%%%%%%%%%%%%%%%%%%%%%%%%%%%%%%%%%%%%%%%%%%%%%%%%%%%%%%%%%%%%%%%%%%%%%%%%%%%%%%

\appendix

\section{Analytical Models}
\label{sec:analytical_model}
If a magnetic flux tube is sufficiently thin so that variations of the physical quantities in the transverse direction of the flux tube can be discarded, and also if the interaction between the flux tube and its surrounding is negligible, the displacement velocity $v_{\parallel}(z,t)$ of a longitudinal wave within the hydrostatically stratified, isothermal flux tube is governed by the Klein-Gordon type equation \citep[][Chapter 8.2]{rae+roberts:1982,hasan+kalkofen:1999,stix:2002}
\begin{equation}
  \frac{\partial^{2}Q_{\parallel}}{\partial t^{2}}
  - c_{\rm T}^{2}\frac{\partial^{2}Q_{\parallel}}{\partial z^{2}} 
  +\omega_{\rm v}^{2}Q_{\parallel} = 0\,.
\label{eqn:Klein-Gordon_longitudinal}
\end{equation}
Here,
\begin{equation}
  Q_{\parallel}(z,t)=e^{-z/4H}v_{\parallel}(z,t)
\label{eqn:formal_solution_longitudinal}
\end{equation}
is the scaled displacement velocity,
\begin{equation}
  c_{\rm T} = \frac{c_{\rm s}c_{\rm A}}{\sqrt{c_{\rm s}^{2} + c_{\rm A}^{2}}}
\label{eqn:tubespeed}
\end{equation}
the tube speed \citep{defouw:1976}, where $c_{\rm s}$ and $c_{\rm A}$ are the speed of sound and the Alfv\'{e}n speed, respectively, and where
\begin{equation}
  \omega_{\rm v}^{2} = \frac{c_{\rm T}^2}{H^2}\left[\frac{1}{16} 
  + \frac{1}{2}(1-\frac{1}{\gamma})(1+\beta)\right]
\label{eqn:cutoff_longitudinal}
\end{equation}
is the cut-off frequency for longitudinal flux-tube waves.
The dispersion relation is then
\begin{equation}
  \omega^{2} = \omega_{\rm v}^{2} + c_{\rm T}^2 k^2\,. 
\label{eqn:dispersion_longitudinal}
\end{equation}
Here, $H=c_{\rm s}^2/\gamma g$ is the pressure scale-height, which is identical to the density scale height for an isothermal atmosphere with constant adiabatic index $\gamma$.
The parameter $\beta=p_{\rm gas}/p_{\rm mag}$ is the ratio of the gas pressure, $p_{\rm gas}$, to the magnetic pressure, $p_{\rm mag}$.
We note that all these equations are also valid for a magnetic flux slab.
In the derivation of Equation~(\ref{eqn:Klein-Gordon_longitudinal}) the difference in the geometry enters via the cross section $A(z)$ of the tube or slab in only the continuity equation, $\partial (A\rho)/\partial t + \partial (A\rho v_{\parallel})/\partial z = 0$, and the magnetic flux conservation, $BA={\rm const.}$, which, however, can be summarized to 
\begin{equation}
  \frac{\partial}{\partial t}\left(\frac{\rho}{B}\right)
  + \frac{\partial}{\partial z}\left(\frac{\rho v_{\parallel}}{B}\right) = 0\,,
\label{eqn:walen}
\end{equation}
where $\rho$ is the mass density and $B$ the magnetic field strength of the thin magnetic flux tube or slab.
Equation~(\ref{eqn:walen}) is a form of Wal\'en's equation in 1D and is independent of the tube/slab cross-section $A(z)$, hence, the independency of  Equation~(\ref{eqn:Klein-Gordon_longitudinal}) on $A(z)$.
Similarly, the displacement $\xi(z,t)$ of a transversal wave within a flux tube is governed by the Klein-Gordon type equation \citep[][Chapter 8.2]{hasan+kalkofen:1999,stix:2002}
\begin{equation}
  \frac{\partial^{2}Q_{\perp}}{\partial t^{2}}
  - c_{\rm kink}^{2}\frac{\partial^{2}Q_{\perp}}{\partial z^{2}} 
  +\omega_{\rm k}^2 Q_{\perp} = 0\,,
\label{eqn:Klein-Gordon_transversal}
\end{equation}
where,
\begin{equation}
  Q_{\perp}(z,t)=e^{-z/4H}{\xi}(z,t)
\label{eqn:formal_solution_transversal}
\end{equation}
is the scaled displacement,
\begin{equation}
  c_{\rm kink} = \sqrt{\frac{2gH}{2\beta + 1}}
\label{eqn:kinkspeed}
\end{equation}
the kink speed, and 
\begin{equation}
  \omega_k^{2} = \frac{g}{8 H}\frac{1}{1+2\beta}
\label{eqn:cutoff_transverse}
\end{equation}
the cut-off frequency for transverse kink waves \citep{spruit:1981,hasan+kalkofen:1999}. 
The dispersion relation is then
\begin{equation}
  \omega^{2} = \omega_k^{2} (16 H^2 k^2 + 1)\,.
\label{eqn:dispersion_longitudinal}
\end{equation}
Also here, the derivation of Equation~(\ref{eqn:Klein-Gordon_transversal})
does not make use of the cross-section $A(z)$ of the thin tube or slab.
Therefore again, all these equations are also valid for a magnetic 
flux slab.

For an isothermal atmosphere, the plasma-$\beta$ as a function of height is constant and assumes the same value for a tube or a slab, given the same field strength at a given, single reference height---a consequence of transverse pressure balance and magnetic flux conservation. 
Therefore, the different expansion rates of a tube vs.\ a slab do not modify the tube speed or the kink speed and neither do they change the cut-off frequencies $\omega_k$ and $\omega_{\rm v}$ so that also the amplitudes of longitudinal and transverse waves are independent of the geometry.
From this it follows that the energy fluxes of sausage and kink waves are independent of whether considering a thin flux tube or a thin flux slab, which justifies the comparison of the present analytical model with our two-dimensional slab model.

\section{Development of shocks in the flux sheath}
\label{sec:shocks}
%%% Describe the formation of shocks in the flux tube

%%% Figure 10: Time evolution of longitudinal wave
\begin{figure}
\epsscale{0.75}
\plotone{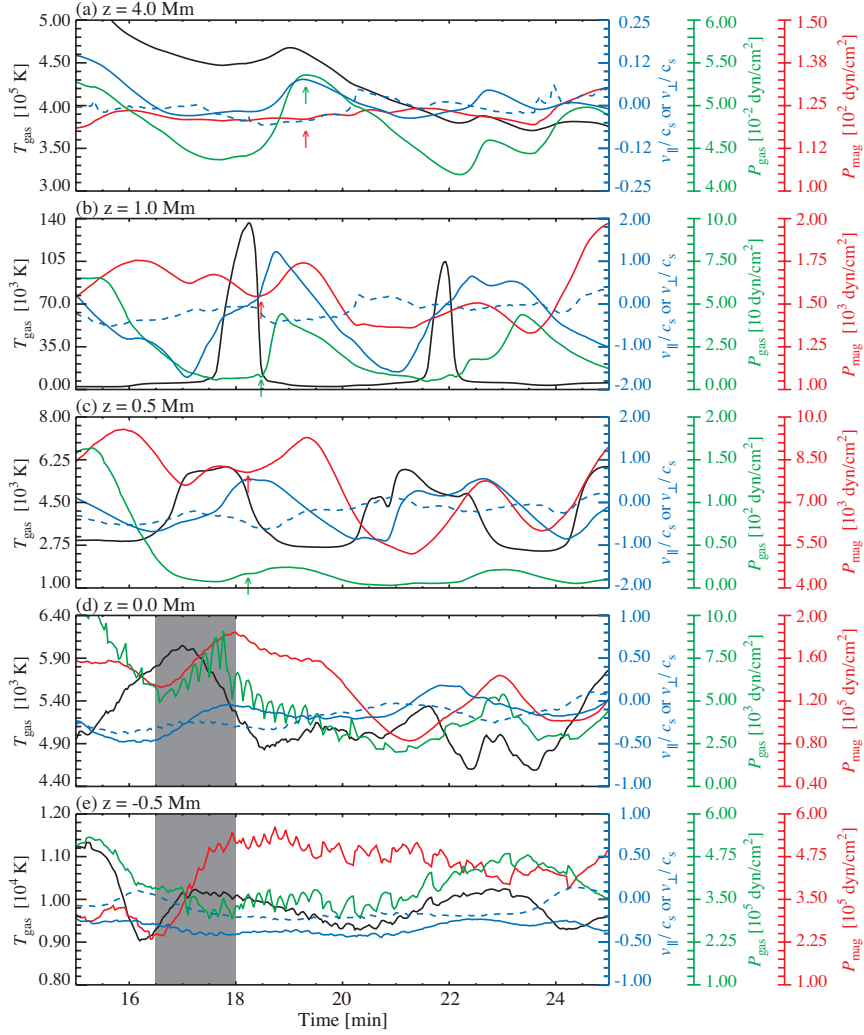}
\caption{Time evolution of atmospheric parameters in the core of the flux slab at different geometrical heights from (a) the lower corona, (b) the lower chromosphere,  (c) the upper photosphere,  (d) near the surface of optical depth unity, to (e) the upper convection zone for $t=15 - 25$ minutes.  Temperature (black solid), gas pressure (green), and magnetic pressure (red) are shown.  Both the longitudinal velocity (blue solid) and the transverse velocity (blue dashed) are normalised to the local sound speed, $c_{\rm s}$.  {\it Shaded region}: An event of large enhancement of the magnetic pressure of more than the time-averaged gas pressure at the surface of optical depth unity ($z\approx 0$).  Red and green arrows indicate a slow mode ($\delta p_{\rm gas} > 0$ and $\delta p_{\rm mag} < 0$) whose vertical propagation speed is $\sim 8.3 - 36~{\rm km\,s^{-1}}$.}
\label{fig:timing1}
\end{figure}

%%% Figure 11: Time evolution of transverse wave
\begin{figure}
\epsscale{0.8}
\plotone{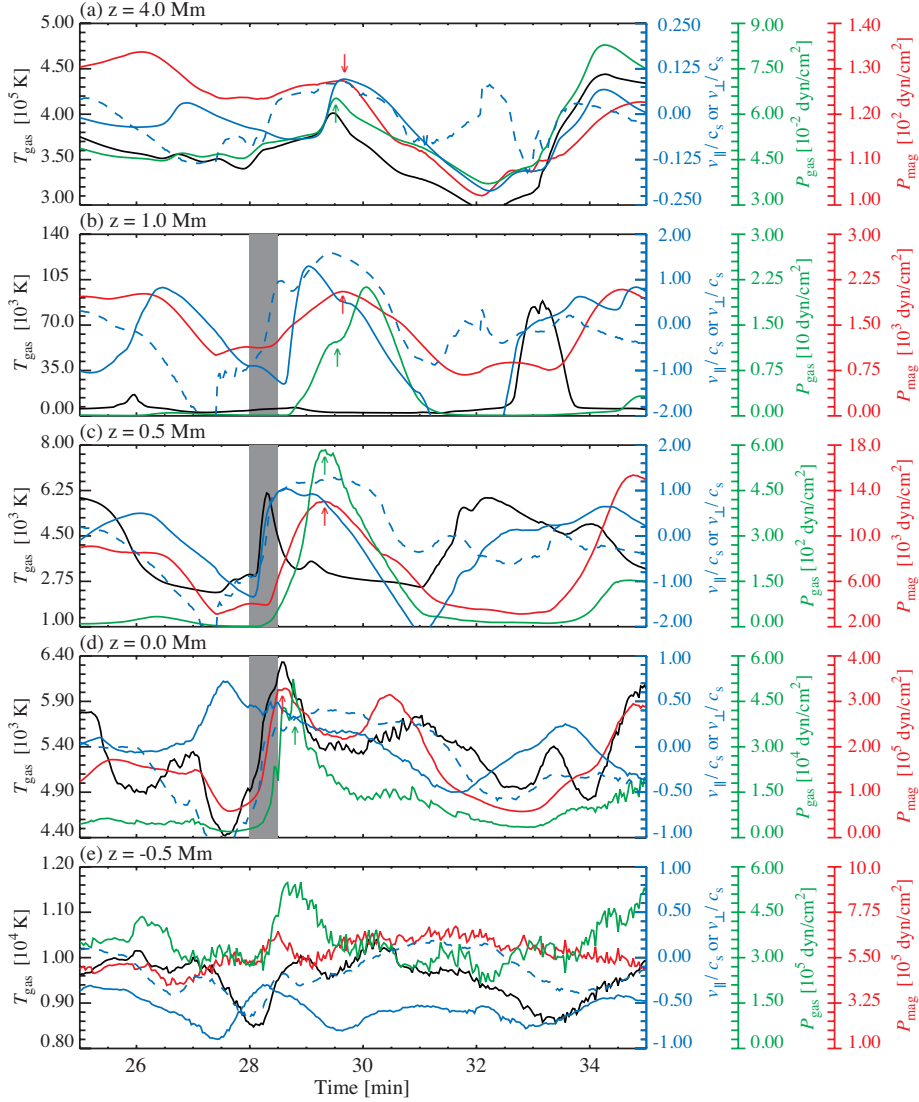}
\caption{Same as Figure\,\ref{fig:timing1}, but for a different time window $t=25 - 35$ minutes.  {\it Shaded region}: An event of rapid and large change of transverse velocity amplitude ($|v_{\perp}/c_{\rm s}|>1$). Red and green arrows indicate a transverse body wave ($\delta p_{\rm gas} > 0$ and $\delta p_{\rm mag} > 0$) whose vertical propagation speed is $\sim 8.3 - 58~{\rm km\,s^{-1}}$.}
\label{fig:timing2}
\end{figure}

%%% Magnetic pumping, slow mode waves, and shocks in the upper atmosphere
Figure\,\ref{fig:timing1} depicts the time evolution of various atmospheric parameters at different geometrical heights in the time period from $t=15$ minutes to $t=25$ minutes.
The shaded regions between $t=16.5$ minutes and $t=18$ minutes at $z=0.0$~Mm and $z=-0.5$~Mm illustrate a large enhancement of magnetic pressure (red curves) after the onset of the first strong magnetic pumping that we already mentioned in Section\,\ref{sec:onset}.\footnote{Gas pressure and magnetic pressure (and to lower degree also the temperature) show an oscillating pattern (of period $< 1$ minute), which is an artifact of the error associated with the tracking of the flux slab of the order of the horizontal grid spacing, $\Delta x$.  It is present only in the surface and subsurface layers on panels (d) and (e) where the width in the core of the flux slab is only a few $\Delta x$ wide.}
In this region, the change in magnetic pressure is larger than or similar to the time-averaged gas pressure, $\Delta p_{\rm mag}\gsim p_{\rm gas}\approx 5\times 10^{3}$ dyn\,cm$^{-2}$ and $\approx 4\times 10^{5}$ dyn\,cm$^{-2}$ in panels (d) and (e), respectively (see also Figure\,\ref{fig:fluxtube-profile}e).
This proves that in the surface layers, the disturbances in the magnetic field strength completely dominate the gas pressure within the flux slab and determine gas pressure fluctuations there.
Directly after the enhancement of the magnetic pressure in the convection zone, the gas pressure (green curves) begins increasing in the photosphere and in the chromosphere consecutively, as indicated by the green arrows at $z=0.5$~Mm and $z=1.0$~Mm.
At the same time, the longitudinal velocity $v_{\parallel}$ (blue solid curves) starts increasing in the photosphere and becomes transonic in the chromosphere in panels (c) and (b), respectively.
The gas temperature (black curves) first increases in panels (c) and (b) when the preceding rarefactional downflow transports hot plasma from higher layers to the respective height levels at $z=0.5$~Mm and $z=1.0$~Mm.  
But later, when the shock passes, it sharply decreases because of photospheric material that is injected into the chromosphere by the following upflow.
In the course of this event, the magnetic pressure appears to be anti-correlated with the gas pressure, namely $\delta p_{\rm mag} < 0$ and $\delta p_{\rm gas} > 0 $ as indicated by red and green arrows in panels (a), (b), and (c).
The shock that develops is best seen in panel (b) but it is not a sharp discontinuity because of the horizontal averaging over the core region of the flux slab.
It dissipates while pushing up the transition zone.
What we see in panel (a) at $z=4.0$~Mm is the mere compressive offspring of this shock.
This magnetic pumping event is strong enough for the entire atmosphere of the flux slab to start oscillating with the cut-off frequency.
After the first magnetic pumping event (shaded region), the longitudinal velocity changes from downward (negative) to upward (positive) and exposes a sawtooth profile in the panels (a), (b), and (c), which is well known from shock-train profiles \citep[Section 5.3]{mihalas+mihalas:1984} and also from the simulation and observation of dynamic fibrils \citep{heggland+:2007, de_pontieu+:2007a}.
%

%%%
Figure\,\ref{fig:timing2} shows the time evolution of the atmospheric parameters during the event of a sudden directional change of the transverse motion of the flux slab just before $t=30$ minutes, as was already mentioned in connection with Figures\,\ref{fig:diameter} and \ref{fig:propagation}.
The shaded regions between $t=28.0$ minutes and $t=28.5$ minutes mark a rapid and large change of the transverse velocity amplitude of $|\Delta v_{\perp}/c_{\rm s}|\sim 2$ in panels (b), (c), and (d), as well as of the longitudinal velocity amplitude of $|\Delta v_{\parallel}/c_{\rm s}|\sim 2$ in the panels (b) and (c).
Because the transverse velocity change from negative to positive takes place over a wide height range at almost the same time, it is likely the impact of a vertically extending shock wave traveling in the horizontal direction in the photosphere and the chromosphere that causes this event.
Both gas pressure and magnetic pressure in the flux slab sharply increase during the impact ($\leq 30$ s), developing into pressure peaks that start traveling upward as indicated by green and red arrows in panels (a), (b), and (c).
Furthermore, the magnetic pressure change is correlated with the gas pressure change, namely $\delta p_{\rm mag} > 0$ and $\delta p_{\rm gas} > 0 $, which points to a transverse body wave.

\section{Time-dependent wavelet power spectra}
\label{sec:local_power}

%%% Figure 13: Local wavelet power
\begin{figure*}
\epsscale{1.0}
\plotone{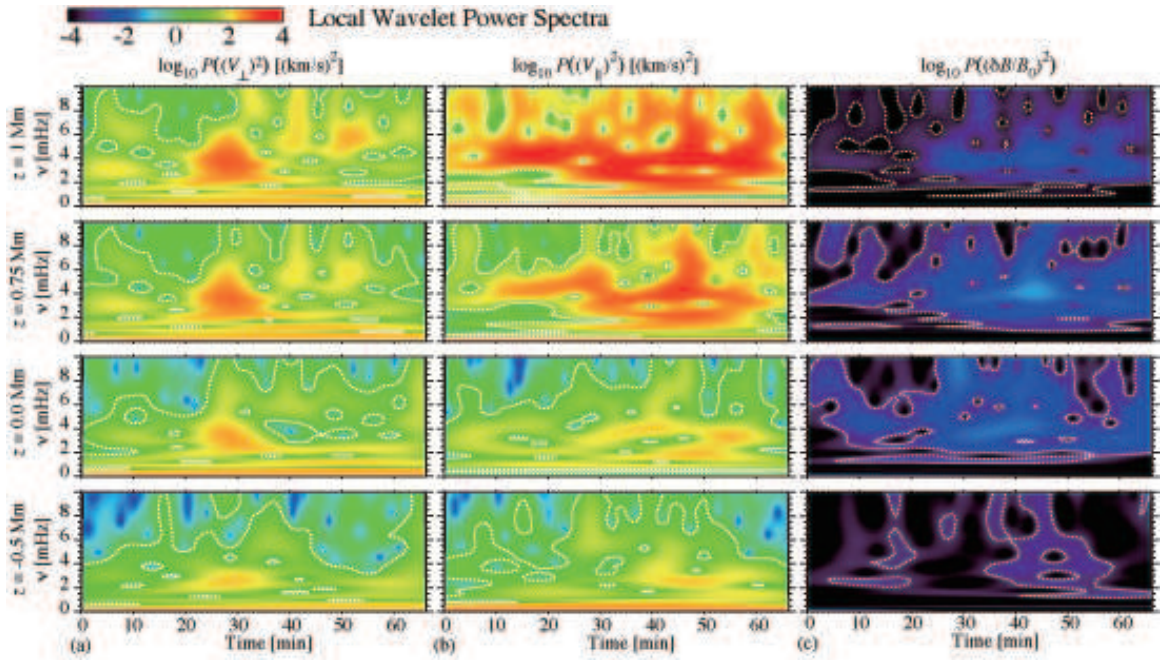}
\caption{Local wavelet power distribution of (a) the transverse velocity,  (b) the longitudinal velocity, and (c) the magnetic flux density fluctuation $\delta B(t)\equiv B(t+\delta t) - B(t)$ normalised by the time-averaged magnetic flux density in the flux-slab core as in Figure\,\ref{fig:global_wavelets}.  The row panels from top to bottom indicate different heights (z=1, 0.75, 0.0, and -0.5~Mm respectively).  The white dotted contour encloses region of greater than $95\%$ confidence level.}
\label{fig:local_wavelets}
\end{figure*}

%% Persistent longitudinal oscillation vs Transient transverse oscillation
Panels (a), (b), and (c) of Figure\,\ref{fig:local_wavelets} show the local wavelet power spectra of the same atmospheric parameters as in panels (a), (b), and (d) of Figure\,\ref{fig:global_wavelets}, respectively, as a function of time at different heights ($z = -0.5$, $0.0$, $0.5$, $1.0$, and $4.0$~Mm).
By comparing panel (a) with panel (b) in Figure\,\ref{fig:local_wavelets} above the surface of optical depth unity ($z\geq 0$~Mm), it becomes obvious that the longitudinal oscillations are persistent whereas the transverse oscillations are transient.
In panel (b) at $z=1$~Mm, the longitudinal oscillations maintain the same frequency $\nu\approx 4$~mHz for the full time period except for the first $10$ minutes and the last $5$ minutes.
This persistence indicates that the longitudinal oscillations near $4$~mHz are regulated by a resonance of the flux-slab atmosphere.
%

%% Transient transverse oscillation at the sporadic event
In panel (a), a transient transverse oscillation ($\nu\approx 3.3$~mHz) is excited at $t\approx 25$ minutes and decays within approximately $10$ minutes at all heights.
At the instant of the excitation of the transverse oscillations, there is a strong transverse motion (from left to right and back again) as can be seen from Figures\,\ref{fig:propagation}\,b and \ref{fig:timing2}.
The maximum local wavelet power of the transverse waves is the same as the global transverse wavelet power, that is $2.3$ in logarithmic scales, meaning $v_{\perp}\approx 14~{\rm km\,s^{-1}}$, which is much larger than the local sound speed.
Such a large velocity disturbance can be caused by a shock propagating in the external atmosphere. 

%% Magnetic oscillations (high frequency oscillations?)
In panel (c), persistent magnetic oscillations are detected at frequencies $\nu\approx 3-4$~mHz in the photosphere as well as in the chromosphere.
By comparing panel (b) with panel (c) of Figure\,\ref{fig:local_wavelets}, one can see that the overall trend in the local power distribution of the magnetic oscillations with confidence level greater than $95\%$ resembles that of the longitudinal oscillations above the surface of optical depth unity.  This finding indicates that the longitudinal oscillations are coupled with the magnetic oscillations, which is what can be expected from magnetic pumping.
%

%%%%%%%%%%%%%%%%%%%%%%%%%%%%%%%%%%%%%%%%%%%%%%%%%%%%%%%%%%%%%%%%%%%%%%%%%%%%%%%%

%%%%%%%%%%%%%%%%%%%%%%%%%%%%%%%%%%%%%%%%%%%%%%%%%%%%%%%%%%%%%%%%%%%%%%%%%%%%%%%%

\bibliography{ms}

\begin{thebibliography}{}
\expandafter\ifx\csname natexlab\endcsname\relax\def\natexlab#1{#1}\fi

\bibitem[{Beckers(1968)}]{beckers:1968}
Beckers, J.~M. 1968, \solphys, 3, 367

\bibitem[{Bel \& Leroy(1977)}]{bel+leroy:1977}
Bel, N., \& Leroy, B. 1977, \aap, 55, 239

\bibitem[{Bello~Gonz{\'a}lez {et~al.}(2010)Bello~Gonz{\'a}lez, Franz,
  Mart{\'\i}nez-Pillet, Bonet, Solanki, del Toro~Iniesta, Schmidt, Gandorfer,
  Domingo, Barthol, Berkefeld, \& Kn{\"o}lker}]{bello_gonzalez+:2010}
Bello~Gonz{\'a}lez, N., Franz, M., Mart{\'\i}nez-Pillet, V., {et~al.} 2010,
  \apjl, 723, L134

\bibitem[{Bogdan {et~al.}(2003)Bogdan, Carlsson, Hansteen, McMurry, Rosenthal,
  Johnson, Petty-Powell, Zita, Stein, McIntosh, \& Nordlund}]{bogdan+:2003}
Bogdan, T.~J., Carlsson, M., Hansteen, V.~H., {et~al.} 2003, \apj, 599, 626

\bibitem[{Cally(2007)}]{cally:2007}
Cally, P.~S. 2007, Astron. Nachr., 328, 286

\bibitem[{Carlsson \& Stein(1997)}]{carlsson+stein:1997}
Carlsson, M., \& Stein, R.~F. 1997, \apj, 481, 1

\bibitem[{Carlsson \& Stein(2002)}]{carlsson+stein:2002}
---. 2002, \apj, 572, 626

\bibitem[{Choudhuri {et~al.}(1993)Choudhuri, Auffret, \&
  Priest}]{choudhuri+:1993a}
Choudhuri, A.~R., Auffret, H., \& Priest, E.~R. 1993, \solphys, 143, 49

\bibitem[{Cranmer \& van Ballegooijen(2005)}]{cranmer+van_ballegooijen:2005}
Cranmer, S.~R., \& van Ballegooijen, A.~A. 2005, ApJS, 156, 265

\bibitem[{De~Pontieu {et~al.}(2004)De~Pontieu, Erd{\'e}lyi, \&
  James}]{de_pontieu+:2004}
De~Pontieu, B., Erd{\'e}lyi, R., \& James, S.~P. 2004, Nature, 430, 536

\bibitem[{De~Pontieu {et~al.}(2007)De~Pontieu, Hansteen, Rouppe van~der Voort,
  van Noort, \& Carlsson}]{de_pontieu+:2007a}
De~Pontieu, B., Hansteen, V.~H., Rouppe van~der Voort, L., van Noort, M., \&
  Carlsson, M. 2007, \apj, 655, 624

\bibitem[{De~Pontieu {et~al.}(2014)De~Pontieu, Rouppe van~der Voort, McIntosh,
  Pereira, Carlsson, Hansteen, Skogsrud, Lemen, Title, Boerner, Hurlburt,
  Tarbell, Wuelser, De~Luca, Golub, McKillop, Reeves, Saar, Testa, Tian,
  Kankelborg, Jaeggli, Kleint, \& Martinez-Sykora}]{de_pontieu+:2014b}
De~Pontieu, B., Rouppe van~der Voort, L., McIntosh, S., {et~al.} 2014, Science,
  346, 1255732

\bibitem[{Defouw(1976)}]{defouw:1976}
Defouw, R.~J. 1976, \apj, 209, 266

\bibitem[{Fedun {et~al.}(2011)Fedun, Shelyag, Verth, Mathioudakis, \&
  Erd{\'e}lyi}]{fedun+:2011}
Fedun, V., Shelyag, S., Verth, G., Mathioudakis, M., \& Erd{\'e}lyi, R. 2011,
  Annales Geophysicae, 29, 1029

\bibitem[{Ferriz-Mas {et~al.}(1989)Ferriz-Mas, Sch{\"u}ssler, \&
  Anton}]{ferriz-mas+:1989}
Ferriz-Mas, A., Sch{\"u}ssler, M., \& Anton, V. 1989, \aap, 210, 425

\bibitem[{Freytag {et~al.}(2012)Freytag, Steffen, Ludwig, Wedemeyer-B{\"o}hm,
  Schaffenberger, \& Steiner}]{freytag+:2012}
Freytag, B., Steffen, M., Ludwig, H.-G.~G., {et~al.} 2012, Journal of
  Computational Physics, 231, 919

\bibitem[{Fujimura \& Tsuneta(2009)}]{fujimura+tsuneta:2009}
Fujimura, D., \& Tsuneta, S. 2009, \apj, 702, 1443

\bibitem[{Giagkiozis {et~al.}(2015)Giagkiozis, Fedun, Erd{\'e}lyi, \&
  Verth}]{giagkiozis+:2015}
Giagkiozis, I., Fedun, V., Erd{\'e}lyi, R., \& Verth, G. 2015, \apj, 810, 53

\bibitem[{Gudiksen {et~al.}(2011)Gudiksen, Carlsson, Hansteen, Hayek,
  Leenaarts, \& Martinez-Sykora}]{gudiksen+:2011}
Gudiksen, B.~V., Carlsson, M., Hansteen, V.~H., {et~al.} 2011, \aap, 531, A154

\bibitem[{Hansteen {et~al.}(2006)Hansteen, De~Pontieu, Rouppe van~der Voort,
  van Noort, \& Carlsson}]{hansteen+:2006}
Hansteen, V.~H., De~Pontieu, B., Rouppe van~der Voort, L., van Noort, M., \&
  Carlsson, M. 2006, \apj, 647, L73

\bibitem[{Hasan \& Kalkofen(1999)}]{hasan+kalkofen:1999}
Hasan, S.~S., \& Kalkofen, W. 1999, \apj, 519, 899

\bibitem[{Hasan {et~al.}(2003)Hasan, Kalkofen, van Ballegooijen, \&
  Ulmschneider}]{hasan+:2003}
Hasan, S.~S., Kalkofen, W., van Ballegooijen, A.~A., \& Ulmschneider, P. 2003,
  \apj, 585, 1138

\bibitem[{Hasan \& Ulmschneider(2004)}]{hasan+ulmschneider:2004a}
Hasan, S.~S., \& Ulmschneider, P. 2004, \aap, 422, 1085

\bibitem[{Hasan \& van Ballegooijen(2008)}]{hasan+van_Ballegooijen:2008}
Hasan, S.~S., \& van Ballegooijen, A.~A. 2008, \apj, 680, 1542

\bibitem[{Heggland {et~al.}(2007)Heggland, De~Pontieu, \&
  Hansteen}]{heggland+:2007}
Heggland, L., De~Pontieu, B., \& Hansteen, V.~H. 2007, \apj, 666, 1277

\bibitem[{Heggland {et~al.}(2011)Heggland, Hansteen, De~Pontieu, \&
  Carlsson}]{heggland+:2011}
Heggland, L., Hansteen, V.~H., De~Pontieu, B., \& Carlsson, M. 2011, \apj, 743,
  142

\bibitem[{Hollweg \& Roberts(1981)}]{hollweg+roberts:1981}
Hollweg, J.~V., \& Roberts, B. 1981, \apj, 250, 398

\bibitem[{Jefferies {et~al.}(2006)Jefferies, McIntosh, Armstrong, Bogdan,
  Cacciani, \& Fleck}]{jefferies+:2006}
Jefferies, S.~M., McIntosh, S.~W., Armstrong, J.~D., {et~al.} 2006, \apj, 648,
  L151

\bibitem[{Judge {et~al.}(2010)Judge, Kn{\"o}lker, Schmidt, \&
  Steiner}]{judge+:2010}
Judge, P.~G., Kn{\"o}lker, M., Schmidt, W., \& Steiner, O. 2010, \apj, 720, 776

\bibitem[{Kato {et~al.}(2011)Kato, Steiner, Steffen, \& Suematsu}]{kato+:2011}
Kato, Y., Steiner, O., Steffen, M., \& Suematsu, Y. 2011, \apjl, 730, L24

\bibitem[{Khomenko {et~al.}(2008)Khomenko, Collados, \&
  Felipe}]{khomenko+:2008}
Khomenko, E., Collados, M., \& Felipe, T. 2008, \solphys, 251, 589

\bibitem[{Kuridze {et~al.}(2012)Kuridze, Morton, Erd{\'e}lyi, Dorrian,
  Mathioudakis, Jess, \& Keenan}]{kuridze+:2012}
Kuridze, D., Morton, R.~J., Erd{\'e}lyi, R., {et~al.} 2012, \apj, 750, 51

\bibitem[{Michalitsanos(1973)}]{michalitsanos:1973}
Michalitsanos, A.~G. 1973, \solphys, 30, 47

\bibitem[{Mihalas \& Weibel~Mihalas(1984)}]{mihalas+mihalas:1984}
Mihalas, D., \& Weibel~Mihalas, B. 1984, Foundations of radiation hydrodynamics
  (New York, Oxford University Press)

\bibitem[{Morton {et~al.}(2014)Morton, Verth, Hillier, \&
  Erd{\'e}lyi}]{morton+:2014}
Morton, R.~J., Verth, G., Hillier, A., \& Erd{\'e}lyi, R. 2014, \apj, 784, 29

\bibitem[{Mumford {et~al.}(2015)Mumford, Fedun, \& Erd{\'e}lyi}]{mumford+:2015}
Mumford, S.~J., Fedun, V., \& Erd{\'e}lyi, R. 2015, \apj, 799, 6

\bibitem[{Murawski {et~al.}(2015)Murawski, Solov'ev, Kra{\'s}kiewicz, \&
  Srivastava}]{murawski+:2015}
Murawski, K., Solov'ev, A., Kra{\'s}kiewicz, J., \& Srivastava, A.~K. 2015,
  \aap, 576, A22

\bibitem[{Murawski \& Zaqarashvili(2010)}]{murawski+zaqarashvili:2010}
Murawski, K., \& Zaqarashvili, T.~V. 2010, \aap, 519, A8

\bibitem[{Parker(1974)}]{parker:1974a}
Parker, E.~N. 1974, \apj, 189, 563

\bibitem[{Rae \& Roberts(1982)}]{rae+roberts:1982}
Rae, I.~C., \& Roberts, B. 1982, \apj, 256, 761

\bibitem[{Roberts \& Webb(1978)}]{roberts+webb:1978}
Roberts, B., \& Webb, A.~R. 1978, \solphys, 56, 5

\bibitem[{Sakurai(1982)}]{sakurai:1982}
Sakurai, T. 1982, \solphys, 76, 301

\bibitem[{Schrijver {et~al.}(1989)Schrijver, Cote, Zwaan, \&
  Saar}]{schrijver+:1989}
Schrijver, C.~J., Cote, J., Zwaan, C., \& Saar, S.~H. 1989, \apj, 337, 964

\bibitem[{Simon \& Leighton(1964)}]{simon+leighton:1964}
Simon, G.~W., \& Leighton, R.~B. 1964, \apj, 140, 1120

\bibitem[{Skumanich {et~al.}(1975)Skumanich, Smythe, \&
  Frazier}]{skumanich+:1975}
Skumanich, A., Smythe, C., \& Frazier, E.~N. 1975, \apj, 200, 747

\bibitem[{Spruit(1981)}]{spruit:1981}
Spruit, H.~C. 1981, \aap, 98, 155

\bibitem[{Stangalini {et~al.}(2015)Stangalini, Giannattasio, \&
  Jafarzadeh}]{stangalini+:2015}
Stangalini, M., Giannattasio, F., \& Jafarzadeh, S. 2015, \aap, 577, A17

\bibitem[{Stangalini {et~al.}(2013)Stangalini, Solanki, Cameron, \&
  Mart{\'\i}nez-Pillet}]{stangalini+:2013a}
Stangalini, M., Solanki, S.~K., Cameron, R., \& Mart{\'\i}nez-Pillet, V. 2013,
  \aap, 554, 115

\bibitem[{Steiner {et~al.}(1998)Steiner, Grossmann-Doerth, Kn{\"o}lker, \&
  Sch{\"u}ssler}]{steiner+:1998}
Steiner, O., Grossmann-Doerth, U., Kn{\"o}lker, M., \& Sch{\"u}ssler, M. 1998,
  \apj, 495, 468

\bibitem[{Steiner \& Pizzo(1989)}]{steiner+pizzo:1989}
Steiner, O., \& Pizzo, V.~J. 1989, \aap, 211, 447

\bibitem[{{Stix}(2002)}]{stix:2002}
{Stix}, M. 2002, {The sun: an introduction} (Springer)

\bibitem[{Torrence \& Compo(1998)}]{torrence+compo:1998}
Torrence, C., \& Compo, G.~P. 1998, Bulletin of the American Meteorological
  Society, 79, 61

\bibitem[{Ulmschneider(1974)}]{ulmschneider:1974}
Ulmschneider, P. 1974, \solphys, 39, 327

\bibitem[{Ulmschneider {et~al.}(1991)Ulmschneider, Z{\"a}hringer, \&
  Musielak}]{ulmschneider+:1991}
Ulmschneider, P., Z{\"a}hringer, K., \& Musielak, Z.~E. 1991, \aap, 241, 625

\bibitem[{Vernazza {et~al.}(1981)Vernazza, Avrett, \& Loeser}]{vernazza+:1981}
Vernazza, J.~E., Avrett, E.~H., \& Loeser, R. 1981, \apjs, 45, 635

\bibitem[{Vigeesh {et~al.}(2012)Vigeesh, Fedun, Hasan, \&
  Erd{\'e}lyi}]{vigeesh+:2012}
Vigeesh, G., Fedun, V., Hasan, S.~S., \& Erd{\'e}lyi, R. 2012, \apj, 755, 18

\bibitem[{Vigeesh {et~al.}(2009)Vigeesh, Hasan, \& Steiner}]{vigeesh+:2009}
Vigeesh, G., Hasan, S.~S., \& Steiner, O. 2009, \aap, 508, 951

\bibitem[{Wedemeyer-B{\"o}hm {et~al.}(2012)Wedemeyer-B{\"o}hm, Scullion,
  Steiner, Rouppe van~der Voort, de~La~Cruz~Rodriguez, Fedun, \&
  Erd{\'e}lyi}]{wedemeyer-bohm+:2012}
Wedemeyer-B{\"o}hm, S., Scullion, E., Steiner, O., {et~al.} 2012, Nature, 486,
  505

\end{thebibliography}

\end{document}